\documentclass[journal]{IEEEtran}

\ifx\pdfoutput\undefined
\usepackage{graphicx} \else
\usepackage[pdftex]{graphicx} \fi
\usepackage{subfigure}
\usepackage{amsmath}
\usepackage{dsfont}
\usepackage{amssymb}
\usepackage{algorithm}
\usepackage{subfigure}
\usepackage{cuted}
\usepackage{epstopdf}
\usepackage{cite}
\usepackage{soul,color,bm}
\usepackage{setspace}
\usepackage{amsbsy}

\usepackage{amsfonts}
\usepackage{calc}
\usepackage{array}
\usepackage{amsthm}
\usepackage{algorithmic}
\usepackage{float}
\usepackage{multirow}
\usepackage{nicefrac}
\usepackage{color,soul}
\usepackage{xcolor}
\usepackage[mathscr]{eucal}
\usepackage{booktabs}
\usepackage{textcomp}
\usepackage{cite}
\usepackage{etoolbox}
\usepackage{tablefootnote}

\newtheorem{lem}{Lemma}
\newtheorem{remark}{Remark}

\usepackage{setspace}
\usepackage{lipsum} 
\usepackage{multicol}
\usepackage{float}
\usepackage{graphicx}

\usepackage{etoolbox}
\makeatletter
\patchcmd{\@begintheorem}{\textit}{\mathbf}{}{}
\makeatother
\ifCLASSINFOpdf
\else
\fi

\begin{document}

\title{Beyond Average-Channel-Based Rate Approximations: UAV Trajectory and Scheduling Optimization With Expected Rate Consideration}

\author{Gitae Park and Kisong Lee,~\IEEEmembership{Senior Member,~IEEE}


\thanks{The authors are with the Department of Information and Communication Engineering, Dongguk University, Seoul 04620, South Korea (e-mail: kslee851105@gmail.com).}

}


\maketitle

\begin{abstract}
This paper investigates the joint optimization of trajectory, user scheduling, and time-slot duration in unmanned aerial vehicle (UAV)-assisted wireless communication systems under minimum expected spectral efficiency (SE) constraints. Unlike most existing studies that approximate the expected SE by substituting the random channel gain with its mean value, thereby evaluating the SE at the average channel realization and overestimating the true expected SE due to Jensen’s inequality, we approximate the expected SE by numerically integrating the SE over the channel distributions. Specifically, instead of relying on average-channel-based approximations, we develop a conservative yet tractable quadrature-based approximation by discretizing the associated cumulative distribution functions. The resulting finite-sum representation explicitly accounts for the probabilistic LoS structure and channel fading effects, while remaining tractable for optimization. Leveraging this lower bound, we formulate a mission completion time minimization problem subject to minimum expected-SE requirements for all ground nodes. The resulting problem is a mixed-integer nonconvex optimization, which is tackled via a penalty-based block coordinate descent framework. The proposed algorithm alternately optimizes the scheduling decisions and the UAV trajectory along with adaptive time-slot durations, and maintains feasibility with respect to the original expected-SE constraints by leveraging successive convex approximation and quadratic transform techniques. Simulation results demonstrate that the proposed method strictly satisfies the minimum expected-SE constraints and achieves a significantly shorter mission completion time than conventional average-channel-based approaches, which are shown to yield infeasible or overly conservative solutions.
\end{abstract}

\begin{IEEEkeywords}
Unmanned aerial vehicles, expected rate, trajectory design, resource allocation, convex optimization.
\end{IEEEkeywords}

\IEEEpeerreviewmaketitle

\section{Introduction}

Recent advances in unmanned aerial vehicle (UAV) platforms have significantly expanded their role in wireless communications, driven by rapid improvements in hardware capability and substantial reductions in deployment cost \cite{Gupta2016, Zeng2016}. These developments have positioned UAVs as a promising architectural component for future wireless networks. Unlike conventional terrestrial base stations, UAVs possess inherent mobility in three-dimensional (3D) space, enabling flexible positioning that can be exploited to adapt network topology according to communication demands. By operating at elevated altitudes, UAVs can mitigate the impact of obstacles such as buildings and terrain, resulting in more favorable air-to-ground channel characteristics and improved link reliability \cite{Lin2018}. This mobility advantage is particularly beneficial in scenarios where terrestrial infrastructure is unavailable, damaged, or economically infeasible. In such environments, UAVs can act as aerial data collectors for ground nodes (GNs) distributed across remote or hard-to-access regions.

The inherent mobility of UAVs has prompted extensive research on the joint optimization of UAV trajectories and communication resources to improve wireless network performance \cite{Namvar19,Shakoor21,Su22,Heo2024-2,Xing23,Zhou24,Wu18,Valiulahi21,Lee18,Heo2024-3,Heo24,Park24,Kim24}. Existing studies primarily focused on different performance objectives, including coverage enhancement \cite{Namvar19,Shakoor21}, spectral efficiency (SE) maximization \cite{Su22,Heo2024-2}, and reliability improvement under adverse channel conditions \cite{Xing23,Zhou24}. Along this line of research, joint optimization of user scheduling, transmit power, and UAV trajectory was investigated to mitigate co-channel interference and improve system throughput \cite{Wu18,Valiulahi21}. Cooperative UAV architectures were also explored, where base-station UAVs coordinated with jammer UAVs to enhance secure communications \cite{Lee18,Heo2024-3}. More recently, UAV-assisted communication systems with GNs equipped with energy-harvesting capabilities attracted growing attention, in which UAV trajectories and communication resources were jointly optimized to support wireless power transfer and data transmission \cite{Heo24,Park24,Kim24}. A common underlying assumption in most of the aforementioned studies is the dominance of line-of-sight (LoS) air-to-ground links, under which simplified channel models were often adopted to maintain tractable optimization. However, such assumptions are difficult to justify in practical deployment scenarios, where signal blockage caused by environmental obstacles is unavoidable. This issue becomes particularly pronounced when UAVs operate at relatively low altitudes, as reduced elevation angles significantly increase the likelihood of non-line-of-sight (NLoS) propagation \cite{Al-Hourani14}.

To address these limitations, probabilistic LoS channel models were proposed to characterize the mixed LoS/NLoS nature of air-to-ground communications by explicitly modeling the dependence of LoS probability on UAV altitude \cite{Al-Hourani14}. Using such models, recent studies have reexamined UAV trajectory and resource allocation problems under more realistic channel conditions \cite{Zeng19,Meng22,Lei25,Pan22,Liu22,You20,Duo20,Luo21,He23,Park26,Kim25}. In particular, some studies focused on UAV trajectory design under energy-related constraints or objectives, jointly accounting for propulsion energy consumption and communication performance \cite{Zeng19,Meng22,Lei25}. Others investigated time-constrained UAV operations, where trajectories and radio resources are optimized to satisfy latency or mission completion requirements \cite{Pan22,Liu22}. In \cite{You20}, UAV-enabled data harvesting was studied using adaptive offline–online trajectory design under time-varying channel conditions, while \cite{Duo20} considered trajectory optimization to enhance robustness against jamming attacks. More recent studies extended these designs to multi-UAV systems, where coordination among multiple UAVs is exploited to improve fairness, interference management, or overall system performance \cite{He23,Kim25}. Probabilistic LoS channel models were also adopted in UAV-enabled wireless-powered networks, where the UAV primarily serves as an energy transmitter and trajectory design is coupled with power transfer strategies to support information delivery \cite{Luo21,Park26}.

Despite these advances under probabilistic LoS channel models, the expected SE in many existing studies \cite{Zeng19,Meng22,Lei25,Pan22,Liu22,You20,Duo20,Luo21,He23,Park26,Kim25} is still approximated by substituting the instantaneous channel with its mean value to simplify the optimization problem. As a result, such average-channel-based approximations systematically overestimate the true expected SE due to Jensen’s inequality. While this practice is computationally convenient, its implications for system reliability have largely been overlooked, particularly in joint UAV trajectory and resource optimization, where inaccuracies in expected-SE characterization can distort the resulting UAV strategy and compromise reliability. These observations motivate the development of a principled expected-SE modeling framework that enables reliable optimization under probabilistic propagation conditions. The contributions of our work can be summarized as follows:
\begin{itemize}
     \item We identify a fundamental limitation of conventional average-channel-based SE approximations, which systematically overestimate expected SEs and may lead to unreliable system designs under stochastic channel conditions. To the best of our knowledge, this work is the first to develop a conservative and tractable expected-SE lower bound based on cumulative distribution function (CDF)-domain discretization and quadrature-based reformulation, enabling reliable constraint enforcement in UAV-assisted communications with probabilistic LoS channels.
     
     \item Leveraging the proposed expected-SE characterization, we formulate a joint optimization problem of UAV trajectory, user scheduling, and time-slot duration, and develop a penalty-based block coordinate descent algorithm. By combining successive convex approximation (SCA) and quadratic transform techniques, the proposed algorithm efficiently handles the coupled nonconvexities and achieves efficient convergence with polynomial-time computational complexity.
     
     \item The results establish that accurate expected-SE modeling is a foundational requirement for reliable UAV communication design under probabilistic LoS channels. By ensuring feasibility at the modeling stage, the proposed formulation defines an optimization region consistent with the original expected-SE constraints, thereby enabling reliable and practically meaningful UAV communication design.
     
\end{itemize}

The remainder of this paper is organized as follows. Section II describes the system model and formulates the mission completion time minimization problem under minimum expected SE constraints. Section III discusses the fundamental limitations of average-channel-based rate approximations and develops a conservative and tractable expected-SE lower-bound formulation. Section IV then presents the proposed joint optimization framework for UAV trajectory, scheduling, and time-slot duration. Section V provides simulation results and discussions, and Section VI concludes the paper.

\section{System Model and Problem Statement}

\begin{figure}[t]
    \centerline{\includegraphics[width=0.9\linewidth]{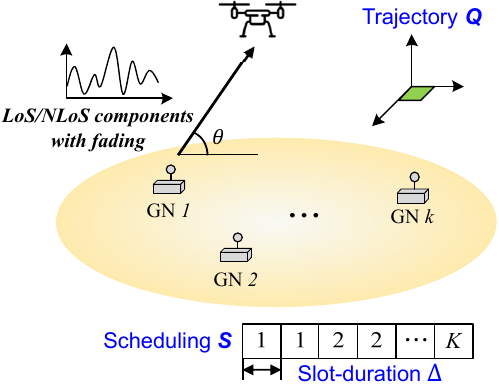}} 
    \caption{System model for a UAV-assisted wireless communication system.}
    \label{fig1}
\end{figure}

\subsection{Constraints on trajectory and communication resources}

As illustrated in Fig. \ref{fig1}, we consider a UAV-assisted wireless communication system in which a rotary-wing UAV collects data from $K$ GNs, indexed by $k \!\in\! \mathcal{K}\!=\!\{1,2,\ldots,K\}$. The UAV mission duration is discretized into $N$ time slots, indexed by $n\!\in\! \mathcal{N}\!=\!\{1,2,\ldots,N\}$, where $\delta[n]$ denotes the duration of time slot $n$. The total mission completion time is thus given by $T\!=\!\sum_{n=1}^{N}\delta[n]$. Assuming sufficiently small time-slot durations, the UAV position is considered constant within each slot \cite{Wu18}, therefore, the slot length is constrained as
\begin{align}
0 \leq \delta[n] \leq \delta_{\textrm{max}}, ~\forall n, \label{const-delta}
\end{align}
where $\delta_{\textrm{max}}$ is chosen sufficiently small such that the wireless channel can be assumed to be flat fading within each time slot. For uplink data transmission, time-division multiple access (TDMA) is employed, such that at most one GN is served in each time slot. 

During time slot $n$, the UAV position is given by $\mathbf{q}[n]\!=\!(x[n],y[n],z[n])$, while GN $k$ remains fixed at $\mathbf{w}_k\!=\!(x_k,y_k,0)$. The UAV is required to travel from $\mathbf{q}_I$ to $\mathbf{q}_F$ while maintaining an operational altitude within $[H_{\textrm{min}}, H_{\textrm{max}}]$ during mission period. Its mobility is constrained by a maximum 3D speed $V_{\textrm{max}}$ and a maximum vertical speed $V_z$, with $V_{\textrm{max}}\ge V_z$ \cite{You20}. Accordingly, the per-slot UAV displacement is limited by $\delta[n]V_{\textrm{max}}$ in space and $\delta[n]V_z$ in altitude. The UAV mobility is subject to the following constraints:
\begin{align}
&\mathbf{q}[0] = \mathbf{q}_I, ~\mathbf{q}[N] = \mathbf{q}_F, \label{constM-1}\\
&H_{\textrm{min}} \leq [\mathbf{q}[n]]_{3} \leq H_{\textrm{max}}, ~\forall n, \label{constM-0}\\ 
&\big\|\mathbf{q}[n]-\mathbf{q}[n\!-\!1]\big\| \leq V_{\textrm{max}}\delta[n], ~ \forall n,  \label{constM-3} \\
&\big\|[\mathbf{q}[n]]_{3}-[\mathbf{q}[n\!-\!1]]_{3}\big\| \leq V_{z}\delta[n], ~ \forall n.  \label{constM-4}
\end{align}

Let $s_k[n]\!\in\!\{0,1\}$ denote the scheduling indicator for GN $k$ at time slot $n$, where $s_k[n]\!=\!1$ indicates that GN $k$ is selected for uplink transmission in slot $n$, and $s_k[n]\!=\!0$ otherwise. Moreover, at most one GN can be scheduled in each time slot, which is expressed as
\begin{align}
&s_{k}[n] \in \{0,1\}, ~\forall k,n, \label{constS-1} \\
&\sum_{k=1}^{K}s_{k}[n] \leq 1, ~\forall n. \label{constS-2} 
\end{align}

\subsection{Probabilistic LoS Channel Model}

To accurately model the air-to-ground channel in UAV communications, we employ a probabilistic LoS channel model, which is particularly suitable for UAV systems. In contrast to fixed base station deployments, this model accounts for the UAV’s ability to increase the probability of LoS links by elevating its altitude, thereby coupling the LoS/NLoS occurrence probabilities with the UAV’s elevation angle.

The LoS probability of the UAV--GN link is determined by an environment-dependent statistical model that captures the effect of building density and the UAV position. Let $c_k[n]\!\in\!\{0,1\}$ denote the corresponding channel state at time slot $n$, where $c_k[n]\!=\!1$ and $c_k[n]\!=\!0$ indicate LoS and NLoS links, respectively. According to the probabilistic LoS model in \cite{Zeng19}, the LoS probability of the channel between the UAV and GN $k$ at time slot $n$ is represented by
\begin{align}
\mathbb{P}(c_k[n]=1) \triangleq P^{\textrm{L}}_{k}[n]=\frac{1}{1+A_1e^{-A_2(\theta_k[n]-A_1)}}, \label{los_prob}
\end{align}
where $A_1$ and $A_2$ are environment-dependent positive constants. The elevation angle between the UAV and GN $k$ at time slot $n$, denoted by $\theta_k[n]\!\in\![0,\frac{\pi}{2}]$, is given by
\begin{align}
\theta_k[n]=\frac{180}{\pi}\arcsin\bigg(\frac{[\mathbf{q}[n]]_{3}}{\|\mathbf{q}[n]-\mathbf{w}_k\|}\bigg), ~\forall k,n. \label{consttheta0} 
\end{align}

Given the LoS probability in \eqref{los_prob}, the corresponding NLoS probability is defined as $\mathbb{P}(c_k[n]\!=\!0)\! \triangleq\! P^{\textrm{N}}_{k}[n]\!=\!1\!-\!P^{\textrm{L}}_{k}[n]$.
Accordingly, the channel power gain between the UAV and GN $k$ at time slot $n$ is modeled as
\begin{align}
h_k[n]=c_k[n]h^{\textrm{L}}_k[n] + (1-c_k[n])h^{\textrm{N}}_k[n], \label{cch}
\end{align}
where $h^{\textrm{L}}_k[n]$ and $h^{\textrm{N}}_k[n]$ denote the channel power gain for the LoS and NLoS conditions, respectively, which are given by
\begin{align}
h^{\textrm{L}}_k[n]&=\frac{|g_k^{\textrm{L}}[n]|^2\beta_{\textrm{L}}}{\|\mathbf{q}[n]\!-\!\mathbf{w}_k\|^{\alpha_{\textrm{L}}}}, \\
h^{\textrm{N}}_k[n]&=\frac{\nu_k[n]|g_{k}^{\textrm{N}}[n]|^{2}\beta_{\textrm{N}}}{\|\mathbf{q}[n]\!-\!\mathbf{w}_k\|^{\alpha_{\textrm{N}}}}.\label{h_nlos}
\end{align}
Here, $(\alpha_{\textrm{L}}, \beta_{\textrm{L}})$ and $(\alpha_{\textrm{N}}, \beta_{\textrm{N}})$ denote the path-loss exponents and reference channel gains for the LoS and NLoS links, respectively, with $\alpha_{\textrm{L}} \!<\! \alpha_{\textrm{N}}$ and $\beta_{\textrm{L}} \!>\! \beta_{\textrm{N}}$ \cite{You20}. The small-scale fading coefficients under LoS and NLoS conditions are denoted by $g_k^{\textrm{L}}[n]$ and $g_k^{\textrm{N}}[n]$, respectively, while $\nu_k[n]$ represents the large-scale shadowing gain. 

The LoS small-scale fading $g_k^{\textrm{L}}[n]$ is modeled by a Rician distribution as
\begin{align}
g_k^{\textrm{L}}[n]=\sqrt{\frac{K_R}{K_R+1}}e^{j\phi_k[n]}+\sqrt{\frac{1}{K_R+1}}\tilde{g}_k[n], \label{Rician}
\end{align}
where $K_R$ denotes the Rician $K$-factor in linear scale, $\phi_k[n]$ is the LoS phase shift, and $\tilde{g}_k[n]\!\sim\!\mathcal{CN}(0,1)$ represents the scattered component.

In contrast, the NLoS small-scale fading is assumed to follow a standard Rayleigh distribution, i.e., $g_k^{\textrm{N}}[n]\!\sim\!\mathcal{CN}(0,1)$. Moreover, the NLoS channel gain in \eqref{h_nlos} incorporates the shadowing effect through $\nu_k[n]$. To ensure a unit-mean shadowing gain, i.e., $\mathbb{E}[\nu_k[n]]=1$, $\nu_k[n]$ is modeled as a bias-corrected log-normal random variable given by
\begin{align}
\nu_k[n]=10^{\frac{\nu^{\textrm{dB}}_k[n]-\Delta_{\textrm{dB}}}{10}},
\end{align}
where $\nu^{\textrm{dB}}_k[n] \!\sim\! \mathcal{N}(0,\sigma_{\textrm{dB}}^2)$ denotes the shadowing component in the dB scale with standard deviation $\sigma_{\textrm{dB}}$, and $\Delta_{\textrm{dB}}\!=\!\frac{\ln 10}{20}\sigma_{\textrm{dB}}^2$ is the bias-correction term.

\subsection{Problem Formulation}

Given the channel power gain in \eqref{cch}, the uplink SE for GN $k$ at time slot $n$ is 
\begin{align}
r_k[n] &= \log_2\left(1+\frac{P_Sh_{k}[n]}{\sigma^2}\right), \label{rk}
\end{align}
where $P_S$ denotes the constant transmit power of GNs and $\sigma^2$ is the noise power. Since the link state is either LoS or NLoS, $r_k[n]$ can be equivalently written as
\begin{align}
r_k[n]=c_k[n]r_k^{\textrm{L}}[n]+(1-c_k[n])r_k^{\textrm{N}}[n],
\end{align}
where $r_k^{\textrm{L}}[n]\!=\!\log_2\big(1\!+\!\frac{P_Sh_k^{\textrm{L}}[n]}{\sigma^2}\big)$ and $r_k^{\textrm{N}}[n]\!=\!\log_2\big(1\!+\!\frac{P_Sh_k^{\textrm{N}}[n]}{\sigma^2}\big)$.

Under the probabilistic LoS channel model, the expected uplink SE can be expressed as follows \cite{Zeng19}: 
\begin{align}
\mathbb{E}[r_k[n]]\!&=\!P_k^{\textrm{L}}[n]\mathbb{E}_{|g_k^{\textrm{L}}[n]|^2}\!\left[r_k^{\textrm{L}}[n]\right] \nonumber\\
&~~~~~~~~~~~~~+\! P_k^{\textrm{N}}[n]\mathbb{E}_{\nu_k[n],|g_k^{\textrm{N}}[n]|^2}\!\left[r_k^{\textrm{N}}[n]\right], \label{expectedrk} 
\end{align}
where the expectations are taken over the corresponding small-scale fading and shadowing random variables under the LoS and NLoS conditions. The resulting time-averaged uplink SE of GN $k$ is given by
\begin{align}\label{SR2}
R_{k}=\frac{1}{T}\sum_{n=1}^Ns_k[n]\delta[n]\mathbb{E}[r_k[n]], ~\forall k.
\end{align}

Our objective is to minimize the UAV mission completion time $T\!=\!\sum_{n=1}^{N}\delta[n]$ while guaranteeing a minimum required SE $R_{\textrm{min}}$ for each GN. To this end, we jointly optimize the scheduling variables $\mathbf{S}\!\triangleq\!\{s_k[n],\,\forall k,n\}$, the UAV 3D trajectory $\mathbf{Q}\!\triangleq\!\{\mathbf{q}[n],\,\forall n\}$, and the time-slot durations $\boldsymbol{\Delta}\!\triangleq\!\{\delta[n],\,\forall n\}$. The resulting optimization problem is formulated as follows.
\begin{align}
\textbf{(P0):}
\min_{\substack{\mathbf{S},~\mathbf{Q},~\boldsymbol{\Delta}}} &~~~~~~~~T \nonumber\\
\textrm{subject to} &~~~ R_{k} \geq R_{\textrm{min}}, ~\forall k, \label{constrk} \\
&~~~\eqref{const-delta}\!-\!\eqref{constS-2}. \nonumber
\end{align}
Problem $\textbf{(P0)}$ is a mixed-integer nonconvex problem due to the binary scheduling variables $\mathbf{S}$ and the nonconvexity of constraint \eqref{constrk} with respect to the optimization variables. Moreover, the expected SE expression in \eqref{constrk} requires evaluating expectations over random fading and shadowing variables, which leads to integral expressions that do not admit closed-form solutions. As a result, obtaining the globally optimal solution to $\textbf{(P0)}$ is highly challenging.

\section{Beyond Average-Channel-Based Rate Approximations}

This section examines the limitations of average-channel-based approximations for expected SE modeling and introduces a quadrature-based lower-bound reformulation. The proposed approach approximates the expected SE under probabilistic LoS channels via a tractable finite-sum representation that is amenable to optimization.

\subsection{Limitations of Average-Channel-Based Rate Approximations}

In offline optimization, the achievable SE involves an expectation over random channel effects such as small-scale fading and shadowing. However, directly handling the expected SE is challenging, as it generally admits only an integral-form expression without a closed-form representation. To address this difficulty, most existing works \cite{Zeng19,Meng22,Lei25,Pan22,Liu22,You20,Duo20,Luo21,He23,Park26,Kim25} approximate the expected SE by substituting the random channel power gains $h_k^{\textrm{L}}[n]$ and $h_k^{\textrm{N}}[n]$ with their mean values, which yields an average-channel-based SE expression given by
\begin{align}
\bar{r}_k[n]& \triangleq P_k^{\textrm{L}}[n]\log_2\!\left(\!1\!+\!\frac{P_S\mathbb{E}_{|g_k^{\textrm{L}}[n]|^2}\left[h_k^{\textrm{L}}[n]\right]}{\sigma^2}\!\right)+\nonumber\\
&~~~~~~P_k^{\textrm{N}}[n]\log_2\!\left(\!1\!+\!\frac{P_S\mathbb{E}_{\nu_k[n],|g_k^{\textrm{N}}[n]|^2}\left[h_k^{\textrm{N}}[n]\right]}{\sigma^2}\!\right).
\end{align}
Since the SE function is concave with respect to the channel power gains, Jensen’s inequality implies that the resulting average-channel-based SE provides an upper bound on the actual expected SE, i.e.,
\begin{align}
\mathbb{E}\left[r_k[n]\right]\leq \bar{r}_k[n].
\end{align}

As a consequence of Jensen’s inequality, replacing the expected SE in constraint \eqref{constrk} by its average-channel-based approximation yields
\begin{align}
\!\!\frac{1}{T}\!\sum_{n=1}^Ns_k[n]\delta[n]\mathbb{E}[r_k[n]] \leq \frac{1}{T}\!\sum_{n=1}^Ns_k[n]\delta[n]\bar{r}_k[n] \triangleq \bar{R}_k.
\end{align}
To conservatively satisfy the expected-SE constraint in \eqref{constrk}, the expected SE must be replaced by a lower bound. 
However, the average-channel-based SE serves only as an estimate of the actual expected SE and, due to Jensen’s inequality, provides an upper bound on it. As a result, even if the estimated SE satisfies the constraint, i.e., $\bar{R}_k \!\ge\! R_{\textrm{min}}$, the actual expected SE $R_k$ may still fall below $R_{\textrm{min}}$, leading to violations of \eqref{constrk} in practice. This issue is particularly critical for mission completion time minimization problems, such as the one considered in this study, where the optimal solution typically lies near the boundary of the feasible set. Optimizing over such a relaxed superset therefore substantially increases the risk of violating the original expected-SE constraints. Motivated by this observation, we depart from the commonly adopted average-channel-based SE approximation and instead seek a conservative yet tractable reformulation of the actual expected SE. Specifically, our objective is to construct a computable lower bound on $\mathbb{E}[r_k[n]]$ that can be directly embedded into the subsequent optimization framework while faithfully capturing the impact of channel randomness.

\subsection{Quadrature-Based Lower-Bound Reformulation}

To construct such a conservative and tractable lower bound, we first rewrite the expected SE $\mathbb{E}\left[r_k[n]\right]$ in \eqref{expectedrk} into an explicit expectation form as
\begin{align}
&\mathbb{E}\left[r_k[n]\right]=P_k^{\textrm{L}}[n]\mathbb{E}_{|g_k^{\textrm{L}}[n]|^2}\!\left[\log_2\!\left(\!1\!+\!\frac{P_S|g_k^{\textrm{L}}[n]|^2\beta_{\textrm{L}}}{\sigma^2\|\mathbf{q}[n]-\mathbf{w}_k\|^{\alpha_{\textrm{L}}}}\!\right)\!\right]\nonumber\\
&+\!P_k^{\textrm{N}}[n]\mathbb{E}_{\nu_k[n],|g_k^{\textrm{N}}[n]|^2}\!\left[\log_2\!\left(\!1\!+\!\frac{P_S\nu_k[n]|g_k^{\textrm{N}}[n]|^2\beta_{\textrm{N}}}{\sigma^2\|\mathbf{q}[n]-\mathbf{w}_k\|^{\alpha_{\textrm{N}}}}\!\right)\!\right].\label{expectedrk2}
\end{align}

Under LoS conditions, the small-scale fading coefficient follows a Rician distribution, and thus the corresponding power gain $|g_k^{\textrm{L}}[n]|^2$ follows a noncentral chi-square distribution parameterized by the Rician $K$-factor. Under NLoS conditions, the small-scale fading coefficient follows a Rayleigh distribution, such that $|g_k^{\mathrm N}[n]|^2$ is exponentially distributed, while the shadowing gain $\nu_k[n]$ is modeled as a log-normal random variable. Noting that $\mathbb{E}[|g_k^{\textrm{L}}[n]|^2]\!=\!\mathbb{E}[|g_k^{\textrm{N}}[n]|^2]\!=\!\mathbb{E}[\nu_k[n]] \!=\! 1$ and that $\nu_k[n]$ and $g_k^{\textrm{N}}[n]$ are statistically independent, the joint distribution associated with the NLoS term can be factorized. 
Consequently, \eqref{expectedrk2} can be expressed in the following integral form:
\begin{align}
&\mathbb{E}\left[r_k[n]\right]\!=\!P_k^{\textrm{L}}[n]\!\int_0^{\infty}\!\!f_{|g_k^{\textrm{L}}[n]|^2}(x)\log_2\!\left(1\!+\!x\zeta_k^{\textrm{L}}[n]\right)dx ~+\nonumber\\
&P_k^{\textrm{N}}[n]\!\int_0^{\infty}\!\!\!\int_0^{\infty}\!\!\!f_{|g_k^{\textrm{N}}[n]|^2}(y)f_{\nu_k[n]}(z)\log_2\!\left(1\!+\!yz\zeta_k^{\textrm{N}}[n]\right)dydz, \label{integral}
\end{align}
where $\zeta_k^{\textrm{L}}[n] \!=\! \frac{P_S\beta_{\textrm{L}}}{\sigma^2\|\mathbf{q}[n]-\mathbf{w}_k\|^{\alpha_{\textrm{L}}}}$ and $\zeta_k^{\textrm{N}}[n]\!=\!\frac{P_S\beta_{\textrm{N}}}{\sigma^2\|\mathbf{q}[n]-\mathbf{w}_k\|^{\alpha_{\textrm{N}}}}$.
Moreover, $f_{|g_k^{\textrm{L}}[n]|^2}(\cdot)$, $f_{|g_k^{\textrm{N}}[n]|^2}(\cdot)$, and $f_{\nu_k[n]}(\cdot)$ denote the probability density functions (PDFs) of $|g_k^{\textrm{L}}[n]|^2$, $|g_k^{\textrm{N}}[n]|^2$, and $\nu_k[n]$, respectively. It is worth noting that the integral expression in \eqref{integral} does not admit a closed-form solution in general, which makes it difficult to incorporate directly into the optimization problem. To overcome this challenge, we employ a numerical quadrature method to derive a tractable finite-sum lower bound for \eqref{integral}, which can be efficiently evaluated and embedded into the proposed optimization framework.

To obtain a tractable finite-sum representation of the expected SE, we discretize the random channel components in both LoS and NLoS links. We first describe the discretization procedure for the LoS small-scale fading term $|g_k^{\textrm{L}}[n]|^2$. The expected SE involves an integral over the PDF of $|g_k^{\textrm{L}}[n]|^2$. A straightforward approach would be to uniformly partition the $|g_k^{\textrm{L}}[n]|^2$ domain along the $x$-axis and approximate the integral accordingly. However, such a uniform partition may lead to large approximation errors, since the probability density of $|g_k^{\textrm{L}}[n]|^2$ can vary significantly across different regions. To address this issue, we adopt a CDF-domain discretization strategy. Specifically, instead of uniformly partitioning the $|g_k^{\textrm{L}}[n]|^2$ domain along the $x$-axis, we uniformly partition the CDF range $[0,1]$ of $|g_k^{\textrm{L}}[n]|^2$ into $U_L$ sub-intervals. This approach ensures that each interval contains an equal probability mass, thereby allocating finer resolution to regions where the fading realizations are more likely to occur. Let $u\!\in\!\{1,\ldots,U_L\}$ index the CDF intervals, and define the $u$-th interval as $\big[\frac{u-1}{U_L},\frac{u}{U_L}\big]$. The representative fading power at the left endpoint of each interval is obtained via the inverse CDF of $|g_k^{\textrm{L}}[n]|^2$, given by
\begin{align}
\gamma^{\textrm{L}}_{u} \triangleq F_{|g_k^{\textrm{L}}[n]|^2}^{-1}\!\left(\frac{u-1}{U_L}\right), ~\forall u, \label{U1}
\end{align}
where $F_{|g_k^{\textrm{L}}[n]|^2}^{-1}(\cdot)$ denotes the inverse CDF of the noncentral chi-square distribution associated with the Rician fading power gain. Since the SE function is monotonically increasing with respect to the fading power gain $|g_k^{\textrm{L}}[n]|^2$, the SE evaluated at $\gamma^{\textrm{L}}_{u}$ serves as a lower bound for all fading realizations within the corresponding CDF interval. As a result, approximating the expected SE by evaluating the SE function at these left-endpoint quantiles yields a conservative finite-sum lower bound:
\begin{align}
&\int_0^{\infty}\!\!f_{|g_k^{\textrm{L}}[n]|^2}(x)\log_2\!\left(1\!+\!x\zeta_k^{\textrm{L}}[n]\right)dx \nonumber\\
&~~~~~~~~~~~~~~~\geq\frac{1}{U_L}\sum_{u=1}^{U_L}\log_2\left(1+\gamma^{\textrm{L}}_{u}\zeta_k^{\textrm{L}}[n]\right).
\end{align}

With this CDF-domain partitioning, we next extend the same discretization principle to the NLoS case, which involves both the small-scale fading power gain $|g_k^{\textrm{N}}[n]|^2$ and the shadowing gain $\nu_k[n]$. Specifically, we uniformly partition the CDF range $[0,1]$ of $|g_k^{\textrm{N}}[n]|^2$ into $U_N$ sub-intervals and that of $\nu_k[n]$ into $U_{\nu}$ sub-intervals. Let $i\!\in\!\{1,\ldots,U_N\}$ and $j\!\in\!\{1,\ldots,U_{\nu}\}$ index the corresponding CDF intervals, defined as $\big[\frac{i-1}{U_N},\frac{i}{U_N}\big]$ and $\big[\frac{j-1}{U_{\nu}},\frac{j}{U_{\nu}}\big]$, respectively. The representative sample points associated with the left endpoints of these CDF intervals are then obtained via the inverse CDFs as
\begin{align}
\gamma^{\textrm{N}}_{i} &\triangleq F_{|g_k^{\textrm{N}}[n]|^2}^{-1}\left(\frac{i-1}{U_N}\right),~\forall i, \label{U2}\\
\gamma^{\nu}_{j} &\triangleq F_{\nu_k[n]}^{-1}\left(\frac{j-1}{U_{\nu}}\right),~\forall j, \label{U3}
\end{align}
where $F_{|g_k^{\textrm{N}}[n]|^2}^{-1}(\cdot)$ and $F_{\nu_k[n]}^{-1}(\cdot)$ represent the inverse CDF of the NLoS small-scale fading power gain, which follows an exponential distribution, and the shadowing gain, which follows a log-normal distribution, respectively.

Using the discretized sample points defined in \eqref{U1}, \eqref{U2}, and \eqref{U3}, we construct a computable lower bound on the expected SE of GN $k$ as follows.
\begin{align}
\underline{r}_k[n] = P_k^{\textrm{L}}[n]\underline{r}_k^{\textrm{L}}[n]+P_k^{\textrm{N}}[n]\underline{r}_k^{\textrm{N}}[n], \label{rkn}
\end{align}
where $\underline{r}_k^{\textrm{L}}[n]$ and $\underline{r}_k^{\textrm{N}}[n]$ are given by
\begin{align}
&\underline{r}_k^{\textrm{L}}[n]\triangleq \frac{1}{U_L}\sum_{u=1}^{U_L}\log_2\!\left(\!1\!+\!\frac{P_S\beta_{\textrm{L}}\gamma^{\textrm{L}}_{u}}{\sigma^2\|\mathbf{q}[n]\!-\!\mathbf{w}_k\|^{\alpha_{\textrm{L}}}}\!\right),\\
&\underline{r}_k^{\textrm{N}}[n]\triangleq\frac{1}{U_N}\frac{1}{U_{\nu}}\sum_{i=1}^{U_N}\sum_{j=1}^{U_{\nu}}\log_2\!\left(\!1\!+\!\frac{P_S\beta_{\textrm{N}}\gamma^{\textrm{N}}_{i}\gamma^{\nu}_{j}}{\sigma^2\|\mathbf{q}[n]\!-\!\mathbf{w}_k\|^{\alpha_{\textrm{N}}}}\!\right). \label{rk_hat}
\end{align}

With the tractable lower bound in \eqref{rkn}, we reformulate problem $\textbf{(P0)}$ by replacing the expected SE $\mathbb{E}[r_k[n]]$ in constraint \eqref{constrk} with its conservative lower bound $\underline{r}_k[n]$, which yields the following problem:
\begin{align}
\textbf{(P1):}
\min_{\substack{\mathbf{S},~\mathbf{Q},~\boldsymbol{\Delta}}} &~~~~~~~~~~~~~~~~~~~~T \nonumber\\
\textrm{subject to} &~~~ \frac{1}{T}\sum_{n=1}^Ns_k[n]\delta[n]\underline{r}_k[n] \geq R_{\textrm{min}}, ~\forall k, \label{constrk2} \\
&~~~\eqref{const-delta}\!-\!\eqref{constS-2}. \nonumber
\end{align}

\section{Proposed Optimization Framework}

Although \textbf{(P1)} admits a tractable expected-SE representation, it is still challenging to solve directly, as feasibility with respect to the minimum-SE constraint \eqref{constrk2} must be ensured from the initial stage of the optimization. In general, finding an initial point $(\mathbf{S}, \mathbf{Q}, \boldsymbol{\Delta})$ that satisfies the minimum-SE constraint for all GNs is difficult, and naive initializations often lead to constraint violations. To address this issue, we adopt a penalty convex--concave procedure by introducing a non-negative slack variable $\rho \ge 0$, which relaxes the minimum-SE constraint during the early iterations. Specifically, constraint \eqref{constrk2} is modified as
\begin{align}
\frac{1}{T}\sum_{n=1}^Ns_k[n]\delta[n]\underline{r}_k[n] \geq R_{\textrm{min}}-\rho, ~\forall k. \label{constrk3}
\end{align}

By penalizing the slack variable $\rho$ in the objective function, problem \textbf{(P1)} is reformulated into the following penalized problem:
\begin{align}
\textbf{(P2):}
\min_{\substack{\mathbf{S},~\mathbf{Q},~\mathbf{\Delta},~\rho\geq 0}} &~~~~~T+\eta\rho \nonumber\\
\textrm{subject to} ~~~~&~~~\eqref{const-delta}\!-\!\eqref{constS-2},~\eqref{constrk3}. \nonumber
\end{align}
Here, the slack variable $\rho$ explicitly quantifies the violation of \eqref{constrk2} through the relaxed constraint \eqref{constrk3}. The parameter $\eta \!>\! 0$ is a penalty weight that controls the impact of the constraint-violation slack on the objective value. By incorporating the penalty term $\eta\rho$ into the objective function, any violation of the minimum-SE constraint is penalized in proportion to $\rho$. For small values of $\eta$, the optimization is allowed to temporarily violate the original minimum-SE constraint, which facilitates the construction of an initial feasible solution. As $\eta$ increases, the penalty on $\rho$ becomes more severe, driving $\rho$ toward zero and thereby recovering feasibility with respect to the original minimum-SE constraint.

Problem \textbf{(P2)} remains challenging to solve due to its inherent nonconvexity. To address this issue, we decompose \textbf{(P2)} into two subproblems with respect to the scheduling variables and the joint UAV trajectory and time-slot duration variables, respectively. Each subproblem is then transformed into a convex optimization problem using appropriate convexification techniques. These subproblems can be efficiently solved using standard convex optimization solvers such as CVX \cite{Grant}, and are alternately updated via a block coordinate descent framework, as detailed in the following subsections.

\subsection{Scheduling Optimization}

By relaxing the binary scheduling variables $s_k[n]$ to the continuous interval $[0,1]$, the scheduling subproblem for fixed $\mathbf{Q}$ and $\boldsymbol{\Delta}$ can be expressed as
\begin{align}
\textbf{(SP1):} \min_{\mathbf{S},~\rho\geq 0} &~~~~~~   T+\eta\rho  \nonumber \\
\textrm{subject to} &~~~~ 0 \leq s_{k}[n] \leq 1, ~\forall k,n, \\ 
&~~~~ \eqref{constS-2},~\eqref{constrk3}.\nonumber
\end{align}
Problem \textbf{(SP1)} is a linear program and can be efficiently solved using CVX. After obtaining the relaxed solution, a binary scheduling policy can be recovered using the approach in \cite{Wu18}, without loss of optimality.

\subsection{Trajectory and Time-Slot Duration Optimization}

For fixed $\mathbf{S}$, the subproblem of jointly optimizing the UAV trajectory $\mathbf{Q}$ and the time-slot duration $\boldsymbol{\Delta}$ is formulated as
\begin{align}
\textbf{(SP2):} \min_{\mathbf{Q},~\mathbf{\Delta},~\rho\geq 0} &~~~~~~~   T+\eta\rho  \nonumber \\
\textrm{subject to} ~~&~~~~ \eqref{const-delta}-\eqref{constM-4},~\eqref{constrk3}.\nonumber
\end{align}
Problem \textbf{(SP2)} is challenging to solve due to the nonconvexity of the minimum-SE constraint \eqref{constrk3}.

In particular, one of the main difficulties arises from the LoS probability $P_k^{\textrm{L}}[n]$, which is expressed as a sigmoid function of the elevation angle $\theta_k[n]$. Since the elevation angle itself is given by an $\arcsin(\cdot)$ mapping of the UAV position $\mathbf{q}[n]$, $P_k^{\textrm{L}}[n]$ becomes a highly nonconvex function of the trajectory variables, making the associated constraints difficult to handle directly. To decouple this nonlinearity, we introduce an auxiliary variable $\underline{\theta}_k[n]$ that serves as a lower bound on the actual elevation angle, i.e.,
\begin{align}
\underline{\theta}_k[n] \leq \frac{180}{\pi}\arcsin\bigg(\frac{[\mathbf{q}[n]]_{3}}{\|\mathbf{q}[n]-\mathbf{w}_k\|}\bigg), ~\forall k,n. \label{const angle}
\end{align}

Since the LoS probability $P_k^{\textrm{L}}[n]$ is monotonically increasing with respect to the elevation angle, substituting $\underline{\theta}_k[n]$ for $\theta_k[n]$ yields a conservative lower bound:
\begin{align}
P_k^{\textrm{L}}[n] \geq \frac{1}{1+A_1e^{-A_2(\underline{\theta}_k[n]-A_1)}}\triangleq \underline{P}_k^{\textrm{L}}[n].
\end{align}

Moreover, since $\beta_{\textrm{L}} \!>\! \beta_{\textrm{N}}$ and $\alpha_{\textrm{N}} \!>\! \alpha_{\textrm{L}}$, the achievable SE under LoS conditions dominates that under NLoS conditions, i.e., $\underline{r}_k^{\textrm{L}}[n] \!\geq\! \underline{r}_k^{\textrm{N}}[n]$. As a result, the expected SE lower bound $\underline{r}_k[n]$, which is a weighted combination of the LoS and NLoS SEs, is monotonically increasing with respect to $P_k^{\textrm{L}}[n]$. By replacing $P_k^{\textrm{L}}[n]$ with $\underline{P}_k^{\textrm{L}}[n]$ in \eqref{rkn}, we obtain a conservative lower bound of $\underline{r}_k[n]$:
\begin{align}
\underline{r}_k[n] \geq \underline{P}_k^{\textrm{L}}[n]\underline{r}_k^{\textrm{L}}[n]+(1-\underline{P}_k^{\textrm{L}}[n])\underline{r}_k^{\textrm{N}}[n] \triangleq \underline{\hat{r}}_k[n]. \label{hatr}
\end{align}
As such, $\underline{\hat{r}}_k[n]$ resolves the inherent nonconvexity arising from the probabilistic LoS model and its coupling with the UAV trajectory. The resulting expression provides a computationally tractable conservative lower bound on the expected SE, which can be directly embedded into the subsequent trajectory optimization problem.

However, constraint \eqref{const angle}, which is newly introduced by the auxiliary variable $\underline{\theta}_k[n]$, remains difficult to handle due to the presence of the $\arcsin(\cdot)$ function.
To eliminate it, we apply the sine function to both sides, which yields
\begin{align} \label{const angle 2}
\sin\left(\frac{\pi}{180}\underline{\theta}_k[n]\right)\leq \frac{[\mathbf{q}[n]]_{3}}{\|\mathbf{q}[n]-\mathbf{w}_k\|},~\forall k,n.
\end{align}
Since $\sin(\cdot)$ is monotonically increasing over $[0,\frac{\pi}{2}]$, the inequality direction is preserved.

Despite this transformation, constraint \eqref{const angle 2} remains nonconvex. Noting that the left-hand side of \eqref{const angle 2} is concave with respect to $\underline{\theta}_k[n]$, it can be upper-bounded by its first-order Taylor expansion as follows:
\begin{align}
&\sin\left(\frac{\pi}{180}\underline{\theta}_k[n]\right) \leq \sin\left(\frac{\pi}{180}\underline{\theta}^{\textrm{prev}}_k[n]\right) + \nonumber\\
&~~~~~~~\frac{\pi}{180}\cos\left(\frac{\pi}{180}\underline{\theta}^{\textrm{prev}}_k[n]\right)(\underline{\theta}_k[n]-\underline{\theta}^{\textrm{prev}}_k[n]), \label{const angle lb}
\end{align}
where $\underline{\theta}^{\textrm{prev}}_k[n]$ is the value of $\underline{\theta}_k[n]$ at the previous SCA iteration. 

Meanwhile, the right-hand side of \eqref{const angle 2} has a concave--over--convex fractional structure. Under the conditions
$[\mathbf{q}[n]]_3 \!\ge\! 0$ and $\|\mathbf{q}[n]\!-\!\mathbf{w}_k\|\!>\!0$, we apply the quadratic transform \cite{Shen18} to obtain the following concave lower bound:
\begin{align}
&\frac{[\mathbf{q}[n]]_{3}}{\|\mathbf{q}[n]-\mathbf{w}_k\|}\geq 2\lambda_k[n]\sqrt{[\mathbf{q}[n]]_3}-\lambda_k^2[n]\|\mathbf{q}[n]-\mathbf{w}_k\|, \label{QT1}
\end{align}
where $\pmb{\lambda}\!\triangleq\!\{\lambda_k[n],\,\forall k,n\}$ is a set of auxiliary variables.

Combining \eqref{const angle lb} and \eqref{QT1}, constraint \eqref{const angle 2} can be conservatively approximated by the following convex set:
\begin{align}
&\sin\!\left(\frac{\pi}{180}\underline{\theta}^{\textrm{prev}}_k[n]\right) \!+\! \frac{\pi}{180}\cos\!\left(\frac{\pi}{180}\underline{\theta}^{\textrm{prev}}_k[n]\right)(\underline{\theta}_k[n]\!-\!\underline{\theta}^{\textrm{prev}}_k[n]) \nonumber\\
&~~~~~~~~~\leq 2\lambda_k[n]\sqrt{[\mathbf{q}[n]]_3}\!-\!\lambda_k^2[n]\|\mathbf{q}[n]\!-\!\mathbf{w}_k\|,~\forall k,n, \label{const angle 3}
\end{align}
which guarantees the feasibility of the original constraint \eqref{const angle}.

In addition, since the right-hand side of \eqref{const angle 3} is concave in $\lambda_k[n]$, its optimal value is obtained by differentiation as
\begin{align}
\lambda_k^*[n]=\frac{\sqrt{[\mathbf{q}[n]]_3}}{\|\mathbf{q}[n]-\mathbf{w}_k\|},~\forall k,n.\label{lambda-star}
\end{align}

The lower bound $\underline{\hat{r}}_k[n]$ in \eqref{hatr} involves both the LoS probability $\underline{P}_k^{\textrm{L}}[n]$ and its complementary term $1\!-\!\underline{P}_k^{\textrm{L}}[n]$, which results in a coupled and highly nonconvex structure. In particular, while $\underline{P}_k^{\textrm{L}}[n]$ follows a sigmoid form, its complement does not share the same functional structure, which complicates the subsequent convexification. To address this issue, we first establish the following lemma, which allows $1\!-\!\underline{P}_k^{\textrm{L}}[n]$ to be equivalently expressed in a symmetric sigmoid form.

\begin{lem}
For the function $f(x)\!=\!1\!-\!\frac{1}{1+ae^{-b(x-a)}}$, an equivalent expression is given by $f(x)\!=\!\frac{1}{1+ae^{-b(2(a+\ln(a)/b)-x-a)}}$.
\begin{IEEEproof}
Starting from the definition of $f(x)$, we obtain
\begin{align}
f(x)=\frac{1}{1+e^{b(x-a)-\ln a}}.
\end{align}
Let $x_s=a+\frac{\ln a}{b}$, so that $b(x-a)-\ln a=b(x-x_s)$. Hence,
\begin{align}
f(x)=\frac{1}{1+e^{b(x-x_s)}}.
\end{align}
Moreover, using $bx_s=ba+\ln a$, 
\begin{align}
e^{b(x-x_s)}=ae^{-b(2x_s-x-a)}.
\end{align}
Substituting this identity into the denominator yields
\begin{align}
f(x)=\frac{1}{1+ae^{-b(2x_s-x-a)}}, 
\end{align}
which completes the proof.
\end{IEEEproof}
\end{lem}

Using \emph{Lemma 1}, $1\!-\!\underline{P}_k^{\textrm{L}}[n]$ can be expressed in the following form that is symmetric to $\underline{P}_k^{\textrm{L}}[n]$:
\begin{align}
1-\underline{P}_k^{\textrm{L}}[n]=\frac{1}{1+A_1e^{-A_2(2x_s-\underline{\theta}_k[n]-A_1)}},
\end{align}
where $x_s\!=\!A_1\!+\!\frac{\ln(A_1)}{A_2}$. 

This symmetric representation enables both the LoS and NLoS components of $\underline{\hat{r}}_k[n]$ to be expressed in a unified sigmoid-based form with respect to the elevation angle. Therefore, $\underline{\hat{r}}_k[n]$ can be expressed as follows:
\begin{align}
\underline{\hat{r}}_k[n] &= \frac{1}{1\!+\!A_1e^{-A_2(\underline{\theta}_k[n]-A_1)}}\frac{1}{U_L} \nonumber\\
&~~~~~~~~\times \sum_{u=1}^{U_L}\log_2\!\left(\!1\!+\!\frac{P_S\beta_{\textrm{L}}\gamma^{\textrm{L}}_{u}}{\sigma^2\|\mathbf{q}[n]\!-\!\mathbf{w}_k\|^{\alpha_{\textrm{L}}}}\!\right)\nonumber\\
&~~~~~~~~+\frac{1}{1\!+\!A_1e^{-A_2(2x_s-\underline{\theta}_k[n]-A_1)}}\frac{1}{U_NU_{\nu}} \nonumber\\
&~~~~~~~~\times \sum_{i=1}^{U_N}\!\sum_{j=1}^{U_{\nu}}\log_2\!\left(\!1\!+\!\frac{P_S\beta_{\textrm{N}}\gamma^{\textrm{N}}_{i}\gamma^{\nu}_{j}}{\sigma^2\|\mathbf{q}[n]\!-\!\mathbf{w}_k\|^{\alpha_{\textrm{N}}}}\!\right)\!.
\end{align}

To enable a tractable convex approximation of $\underline{\hat{r}}_k[n]$, it is necessary to characterize the convexity of the SE terms with respect to the optimization variables. The following lemma provides a useful convexity result that will be exploited in the subsequent reformulation.

\begin{lem}
For given $c \!\geq \!0$ and $\alpha\!\in\![2,6]$, the function $f(x,y)\!=\!\frac{1}{x}\log_2(1\!+\!\frac{c}{y^{\alpha/2}})$ is convex for $x\!>\!0$ and $y\!>\!0$.
\begin{IEEEproof}
By direct second-order analysis \cite{You20}, it can be verified that for any nonzero vector $\mathbf{t}$, $\mathbf{t}^T \nabla^2 f(x,y) \mathbf{t} \ge 0$ holds for $x\!>\!0$ and $y\!>\!0$ with $c \!\geq \!0$ and $\alpha\!\in\![2,6]$, which establishes the convexity of $f(x,y)$. 
\end{IEEEproof}
\end{lem}

According to \emph{Lemma 2}, $\underline{\hat{r}}_k[n]$ is a convex function with respect to the variables $(1\!+\!A_1e^{-A_2(\underline{\theta}_k[n]-A_1)})$, $(1\!+\!A_1e^{-A_2(2x_s-\underline{\theta}_k[n]-A_1)})$, and $\|\mathbf{q}[n]\!-\!\mathbf{w}_k\|^2$. Therefore, by applying the first-order Taylor approximation of $\underline{\hat{r}}_k[n]$ with respect to these variables, a lower bound of $\underline{\hat{r}}_k[n]$ can be obtained as follows:
\begin{align}
&\underline{\hat{r}}_k[n]\geq \underline{\hat{r}}^{\textrm{prev}}_k[n]\!-\!
\frac{1}{U_L}\!\sum_{u=1}^{U_L}\psi_{k,u}^{\textrm{L},\textrm{prev}}[n](X_{k}^{\textrm{L}}[n]\!-\!X_{k}^{\textrm{L},\textrm{prev}}[n])\nonumber\\
&-\frac{1}{U_NU_{\nu}}\!\sum_{i=1}^{U_N}\sum_{j=1}^{U_{\nu}}\psi_{k,i,j}^{\textrm{N},\textrm{prev}}[n](X_{k}^{\textrm{N}}[n]\!-\!X_{k}^{\textrm{N},\textrm{prev}}[n])\nonumber\\
&-\frac{1}{U_L}\!\sum_{u=1}^{U_L}\chi_{k,u}^{\textrm{L},\textrm{prev}}[n](\|\mathbf{q}[n]\!-\!\mathbf{w}_k\|^2\!-\!\|\mathbf{q}^{\textrm{prev}}[n]\!-\!\mathbf{w}_k\|^2)\nonumber\\
&-\frac{1}{U_NU_{\nu}}\!\sum_{i=1}^{U_N}\sum_{j=1}^{U_{\nu}}\chi_{k,i,j}^{\textrm{N},\textrm{prev}}[n](\|\mathbf{q}[n]\!-\!\mathbf{w}_k\|^2\!-\!\|\mathbf{q}^{\textrm{prev}}[n]\!-\!\mathbf{w}_k\|^2)\nonumber\\
&~~~~~~\triangleq \underline{\hat{r}}^{\textrm{LB}}_k[n], \label{Rlb-q}
\end{align}
where $X_{k}^{\textrm{L}}[n]\!=\!1\!+\!A_1e^{-A_2(\underline{\theta}_k[n]-A_1)}$, $X_k^{\textrm{L},\textrm{prev}}[n]\!=\!1\!+\!A_1e^{-A_2(\underline{\theta}^{\textrm{prev}}_k[n]-A_1)}$, $X_{k}^{\textrm{N}}[n]\!=\!1\!+\!A_1e^{-A_2(2x_s-\underline{\theta}_k[n]-A_1)}$, $X_{k}^{\textrm{N},\textrm{prev}}[n]\!=\!1\!+\!A_1e^{-A_2(2x_s-\underline{\theta}^{\textrm{prev}}_k[n]-A_1)}$, and $\mathbf{q}^{\textrm{prev}}[n]$ represents the UAV trajectory at time slot $n$ for the previous SCA iteration. Moreover, $\underline{\hat{r}}^{\textrm{prev}}_k[n]$, $\psi_{k,u}^{\textrm{L},\textrm{prev}}[n]$, $\psi_{k,i,j}^{\textrm{N},\textrm{prev}}[n]$, $\chi_{k,u}^{\textrm{L},\textrm{prev}}[n]$, and $\chi_{k,i,j}^{\textrm{N},\textrm{prev}}[n]$ are defined as follows:
\begin{align}
&\underline{\hat{r}}^{\textrm{prev}}_k[n]\!=\!\frac{1}{X_k^{\textrm{L},\textrm{prev}}[n]U_L}\!\sum_{u=1}^{U_L}\log_2\!\left(\!1\!+\!\frac{\Gamma^{\textrm{L}}_{u}}{(y_k^{\textrm{prev}}[n])^{\alpha_{\textrm{L}}/2}}\!\right)\nonumber\\
&~~+\!\frac{1}{X_k^{\textrm{N},\textrm{prev}}[n]U_NU_{\nu}}\!\sum_{i=1}^{U_N}\sum_{j=1}^{U_{\nu}}\log_2\!\left(\!1\!+\!\frac{\Gamma^{\textrm{N}}_{i,j}}{(y_k^{\textrm{prev}}[n])^{\alpha_{\textrm{N}}/2}}\!\right), \\
&\psi_{k,u}^{\textrm{L},\textrm{prev}}[n]=\frac{1}{(X_k^{\textrm{L},\textrm{prev}}[n])^2}\log_2\left(1+\frac{\Gamma^{\textrm{L}}_{u}}{(y_k^{\textrm{prev}}[n])^{\alpha_{\textrm{L}}/2}}\right),\\
&\psi_{k,i,j}^{\textrm{N},\textrm{prev}}[n]=\frac{1}{(X_k^{\textrm{N},\textrm{prev}}[n])^2}\log_2\left(1+\frac{\Gamma^{\textrm{N}}_{i,j}}{(y_k^{\textrm{prev}}[n])^{\alpha_{\textrm{N}}/2}}\right),\\
&\chi_{k,u}^{\textrm{L},\textrm{prev}}[n]=\frac{1}{X_k^{\textrm{L},\textrm{prev}}[n]}\frac{\alpha_{\textrm{L}}\Gamma^{\textrm{L}}_{u}\log_2e}{2y_k^{\textrm{prev}}[n]((y_k^{\textrm{prev}}[n])^{\alpha_{\textrm{L}}/2}+\Gamma^{\textrm{L}}_{u})},\\
&\chi_{k,i,j}^{\textrm{N},\textrm{prev}}[n]=\frac{1}{X_k^{\textrm{N},\textrm{prev}}[n]}\frac{\alpha_{\textrm{N}}\Gamma^{\textrm{N}}_{i,j}\log_2e}{2y_k^{\textrm{prev}}[n]((y_k^{\textrm{prev}}[n])^{\alpha_{\textrm{N}}/2}+\Gamma^{\textrm{N}}_{i,j})},
\end{align}
where $\Gamma^{\textrm{L}}_{u}\!=\!\frac{P_S\beta_{\textrm{L}}\gamma^{\textrm{L}}_{u}}{\sigma^2}$, $\Gamma^{\textrm{N}}_{i,j}\!=\!\frac{P_S\beta_{\textrm{N}}\gamma^{\textrm{N}}_{i}\gamma^{\nu}_{j}}{\sigma^2}$, and $y_k^{\textrm{prev}}[n]\!=\!\|\mathbf{q}^{\textrm{prev}}[n]\!-\!\mathbf{w}_k\|^2$.

Using \eqref{Rlb-q}, the minimum-SE constraint \eqref{constrk3} can be conservatively enforced by replacing $\underline{r}_k[n]$ with its lower bound $\underline{\hat{r}}^{\textrm{LB}}_k[n]$, yielding
\begin{align}
\frac{1}{T}\sum_{n=1}^Ns_k[n]\delta[n]\underline{\hat{r}}^{\textrm{LB}}_k[n] \geq R_{\textrm{min}}-\rho, ~\forall k. \label{constR-q}
\end{align}
Constraint \eqref{constR-q} is still nonconvex due to the bilinear term $\delta[n]\underline{\hat{r}}^{\textrm{LB}}_k[n]$. 
To handle this coupling, we rewrite the product in an equivalent concave--over--convex fractional form:
\begin{align}
\delta[n]\underline{\hat{r}}^{\textrm{LB}}_k[n]=\frac{\underline{\hat{r}}^{\textrm{LB}}_k[n]}{1/\delta[n]},
\end{align}
where $\underline{\hat{r}}^{\textrm{LB}}_k[n]$ is concave with respect to the trajectory variables, whereas $1/\delta[n]$ is convex in $\delta[n]$. Moreover, since $\underline{\hat{r}}^{\textrm{LB}}_k[n]$ is nonnegative by construction, applying the quadratic transform only requires $1/\delta[n]\!>\!0$. This condition can be ensured by imposing the mild constraint:
\begin{align}
\delta[n] &\geq \delta_{\textrm{min}}, ~\forall n, \label{const_qt}
\end{align}
where $\delta_{\textrm{min}} \!>\! 0$ is a small positive constant. 

With \eqref{const_qt}, we apply the quadratic transform to $\frac{\underline{\hat{r}}^{\textrm{LB}}_k[n]}{1/\delta[n]}$ and obtain the following concave lower bound of \eqref{constR-q}:
\begin{align}
&\frac{1}{T}\sum_{n=1}^Ns_k[n]\delta[n]\underline{\hat{r}}^{\textrm{LB}}_k[n] \geq \nonumber\\
&\frac{1}{T}\sum_{n=1}^Ns_k[n]\left(2\mu_k[n]\sqrt{\underline{\hat{r}}^{\textrm{LB}}_k[n]}-\frac{\mu_k[n]^2}{\delta[n]}\right)\triangleq \hat{R}_k, \label{rk_qt}
\end{align}
where $\pmb{\mu}\!\triangleq\!\{\mu_k[n],\,\forall k,n\}$ is a set of auxiliary variables.

Using \eqref{rk_qt}, a conservative convex constraint for \eqref{constR-q} is given by
\begin{align}
\hat{R}_k\geq R_{\textrm{min}}-\rho,~\forall k. \label{constR-q2}
\end{align}
It is worth noting that any solution satisfying \eqref{constR-q2} also satisfies \eqref{constR-q}, and hence guarantees the original minimum-SE requirement.

For fixed $(\mathbf{Q},\boldsymbol{\Delta})$, the right-hand side of \eqref{rk_qt} is concave in $\mu_k[n]$, and the optimal auxiliary variable is also obtained in closed form as
\begin{align}
\mu_k^*[n] = \delta[n]\sqrt{\underline{\hat{r}}^{\textrm{LB}}_k[n]},~\forall k,n. \label{omega-star}
\end{align}

Using \eqref{const angle 3}, \eqref{const_qt}, and \eqref{constR-q2}, the nonconvex problem \textbf{(SP2)} can be conservatively approximated by the following convex optimization problem:
\begin{align}
\textbf{(SP2-1):} \min_{\mathbf{Q},~\boldsymbol{\Delta},~\rho\geq0} ~&~~~~~~~~~~~~~~~~~   T+\eta\rho  \nonumber \\
\textrm{subject to} ~~~&~~~~ \eqref{const-delta}-\eqref{constM-4},~\eqref{const angle 3},~\eqref{const_qt},~\eqref{constR-q2}.\nonumber
\end{align}

\subsection{Procedure of Proposed Algorithm}

Both subproblems, \textbf{(SP1)} and \textbf{(SP2-1)}, are convex with respect to their respective optimization variables and can be efficiently solved using standard convex optimization solvers. By alternately solving these two subproblems, the overall algorithm iterates until convergence. The detailed procedure is summarized in Algorithm~\ref{Alg1}.
\begin{algorithm}[h]
    \caption{Proposed Algorithm} \label{Alg1} \small
    1:$~$Set $r\!=\!0$ and initialize $\mathbf{S}^{r}$, $\mathbf{Q}^{r}$, $\boldsymbol{\Delta}^r$, $\eta^r$, $\eta_{\textrm{max}}$, and $\varepsilon > 1$ \\
    2:$~$Calculate $T^r = \sum_{n=1}^{N}\delta[n]$\\
    3:$~$\textbf{repeat}\\
    4:$~~~$Update $r \leftarrow r+1$  \\
    5:$~~~$Update $T^{\textrm{old}} \leftarrow T^{r-1}$\\
    6:$~~~$Find $\mathbf{S}^r$ by solving \textbf{(SP1)} for given $\{\mathbf{Q}^{r-1},\boldsymbol{\Delta}^{r-1}\}$\\
    7:$~~~$Update $\{\pmb{\lambda}^r,\pmb{\mu}^r\}$ using \eqref{lambda-star} and \eqref{omega-star} \\
    8:$~~~$Find $\{\mathbf{Q}^r,\boldsymbol{\Delta}^r\}$ by solving \textbf{(SP2-1)} for given $\mathbf{S}^r$ \\
    9:$~~~$Update $\eta^r \leftarrow \min\{\varepsilon\eta^{r-1},\eta_{\textrm{max}}\}$ \\
    10:$~~$Calculate $T^r=\sum_{n=1}^N\delta[n]$ \\
    11:$~$\textbf{until} $|T^r-T^{\textrm{old}}|<\epsilon$
\end{algorithm}

\begin{remark}[Convergence and Computational Complexity] \label{Remark1}
Algorithm~\ref{Alg1} starts from an initial feasible solution $\{\mathbf{S}, \mathbf{Q}, \boldsymbol{\Delta}\}$ and employs a penalty parameter $\eta$, which is gradually increased by a factor $\varepsilon \!>\! 1$ until it reaches a prescribed upper bound $\eta_{\textrm{max}}$. As established in \cite{Vu16}, there exists a finite $\eta_{\textrm{max}}$ for which the penalty term converges to zero. After $\eta$ attains this upper bound, the objective value does not increase over successive iterations, i.e.,
\begin{align}
T(\mathbf{S}^{r-1}\!,\mathbf{Q}^{r-1}\!, \boldsymbol{\Delta}^{r-1}) \!\geq\! T(\mathbf{S}^r,\mathbf{Q}^r, \boldsymbol{\Delta}^r). 
\end{align}
Moreover, since the objective is lower bounded by a finite constant \cite{Bertsekas99}, the proposed algorithm is guaranteed to converge.

The complexity of the proposed algorithm is analyzed under the standard worst-case framework for interior-point methods \cite{Ben-Tal01,Boyd04}. For a problem with $N_V$ variables, the computational cost of each interior-point iteration scales on the order of $\mathcal{O}(N_V^3)$, while the total number of iterations scales on the order of $\mathcal{O}\left(\sqrt{N_V}\log(1/\epsilon)\right)$, where $\epsilon \!>\!0$ denotes the desired solution accuracy. Taking into account the iterative structure of the proposed algorithm, the resulting worst-case complexity is given by $\mathcal{O}\!\left(R_C (KN)^{3.5}\log(1/\epsilon)\right)$, where $R_C$ is the number of iterations required for convergence (corresponding to lines 3--11 in Algorithm \ref{Alg1}). The resulting polynomial-time complexity suggests that the proposed approach is computationally tractable for practical system sizes \cite{Leiserson}. Note that the discretization parameters $\{U_L,U_N,U_\nu\}$ determine the numerical accuracy of the expected-SE approximation and affect the per-iteration computational cost through the evaluation of constraint \eqref{constR-q2}, but they do not increase the number of optimization variables. Consequently, they appear only as constant factors in the complexity analysis.
\end{remark}

\section{Simulation Results and Discussions}
\begin{table}[ht]
\begin{center}
\caption{Parameter Setup} \footnotesize
\begin{tabular}{ll} \hline 
Description & Value \\ \hline \hline
Number of GNs & $K$ = 4 \\
Number of time slots & $N$ = 160 \\
Max/Min slot lengths & $\{\delta_{\textrm{max}},\delta_{\textrm{min}}\}$ = $\{0.5,10^{-5}\}$ s\\
Max/Min altitudes & $\{H_{\textrm{max}},H_{\textrm{min}}\}$ = $\{200,10\}$ m \\
Max 3D/vertical speeds & $\{V_{\textrm{max}},V_{z}\}$ =  $\{20,\frac{V_{\textrm{max}}}{2}\}$ m/s\\
Transmit power of GNs & $P_S$ =  30 dBm\\
Constants for LoS probability & $\{A_1,A2\} = \{12.08, 0.114\}$\\
Path-loss exponents & $\{\alpha_{\textrm{L}},\alpha_{\textrm{N}}\}$ = $\{2,2.7\}$ \\
Reference channel gains & $\{\beta_{\textrm{L}},\beta_{\textrm{N}}\}$ = $\{-30,-40\}$ dB \\
Noise power & $\sigma^{2}$ = $-70$ dBm \\
Minimum required SE & $R_{\textrm{min}}$ = 2.4 bps/Hz\\
Rician K-factor & $K_{R}$ = 15 dB\\
Shadowing standard deviation & $\sigma_{\textrm{dB}}$ = 10 dB\\
Discretization parameters & $U_L=U_N=U_{\nu}$ = 40\\
Penalty parameters & $\{\eta^0,\eta_{\textrm{max}},\varepsilon\}$ = $\{1,10^5,1.5\}$\\
Convergence threshold & $\epsilon$ = 10$^{-3}$ \\
Number of Monte Carlo realizations & 30000\\
\hline
\end{tabular} 
\label{table1}
\end{center}
\end{table}

\begin{figure*}[ht!]
  \begin{center}
    \subfigure[3D trajectory of the proposed scheme.]{
      \includegraphics[width=0.315\linewidth]{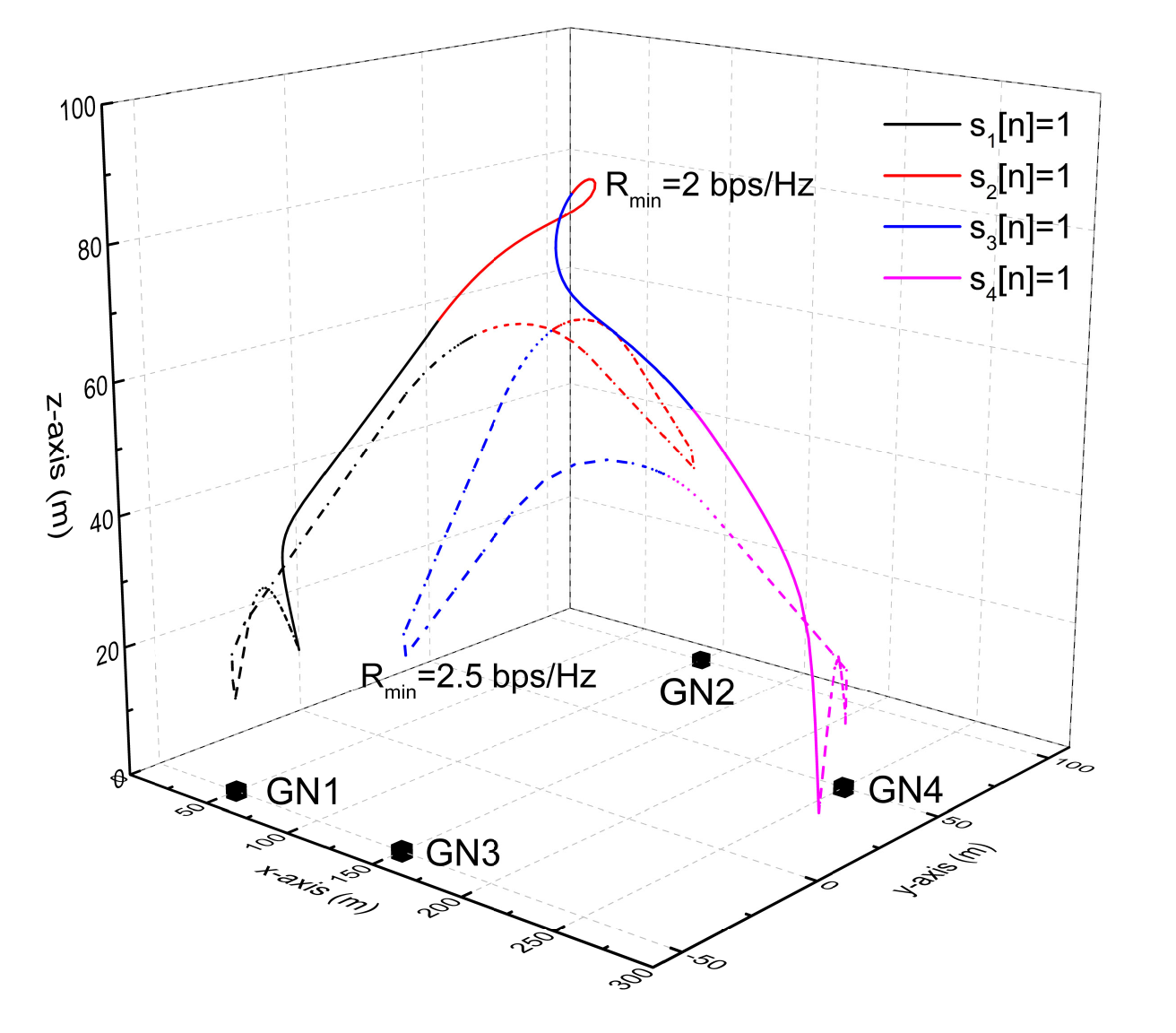}\label{R1-1}
    }
    \subfigure[Horizontal trajectory of the proposed scheme.]{
      \includegraphics[width=0.315\linewidth]{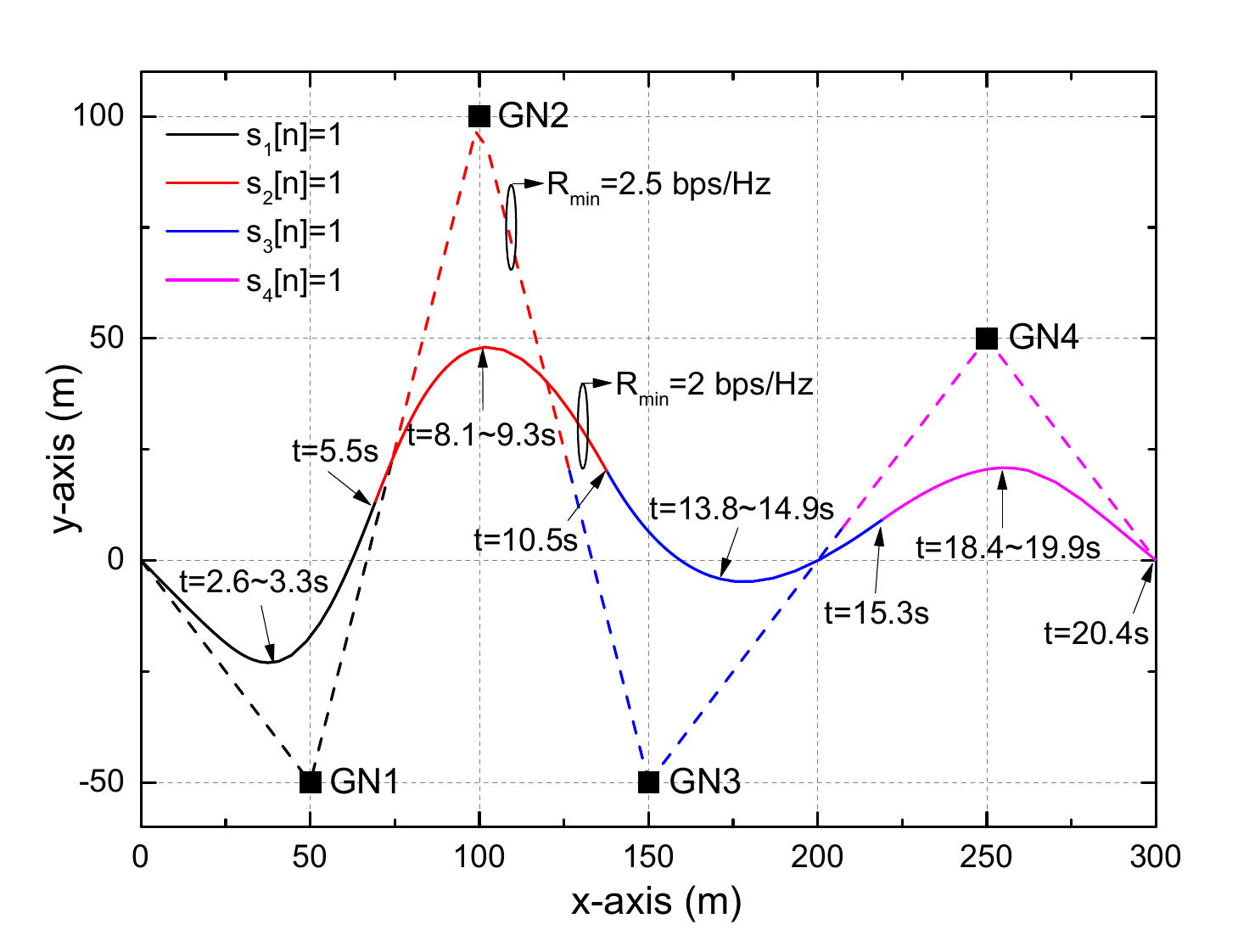}\label{R1-2}
    }
    \subfigure[3D trajectory of the AC-based scheme.]{
      \includegraphics[width=0.315\linewidth]{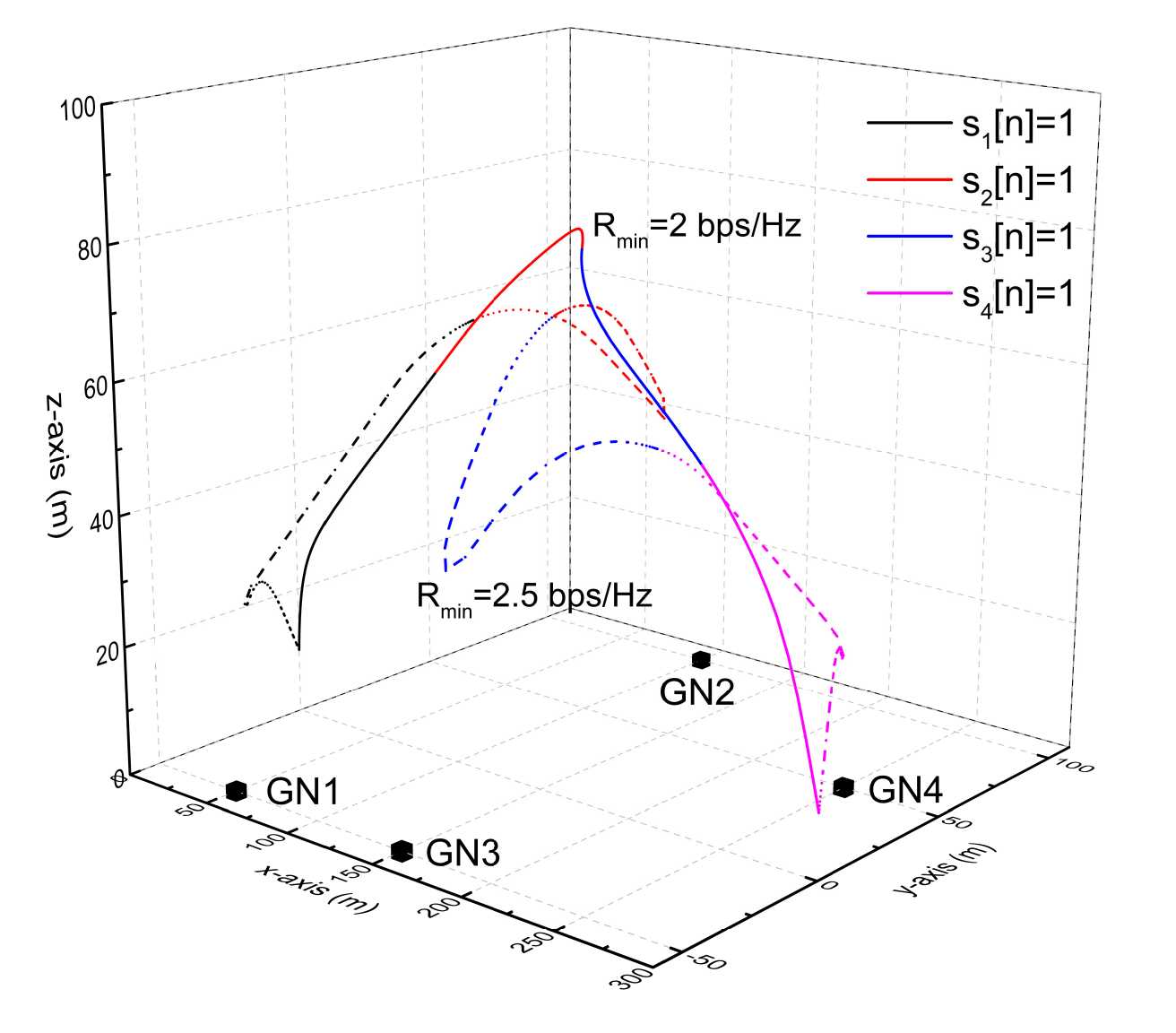}\label{R1-3}
    }
    \subfigure[Horizontal trajectory of the AC-based scheme.]{
      \includegraphics[width=0.315\linewidth]{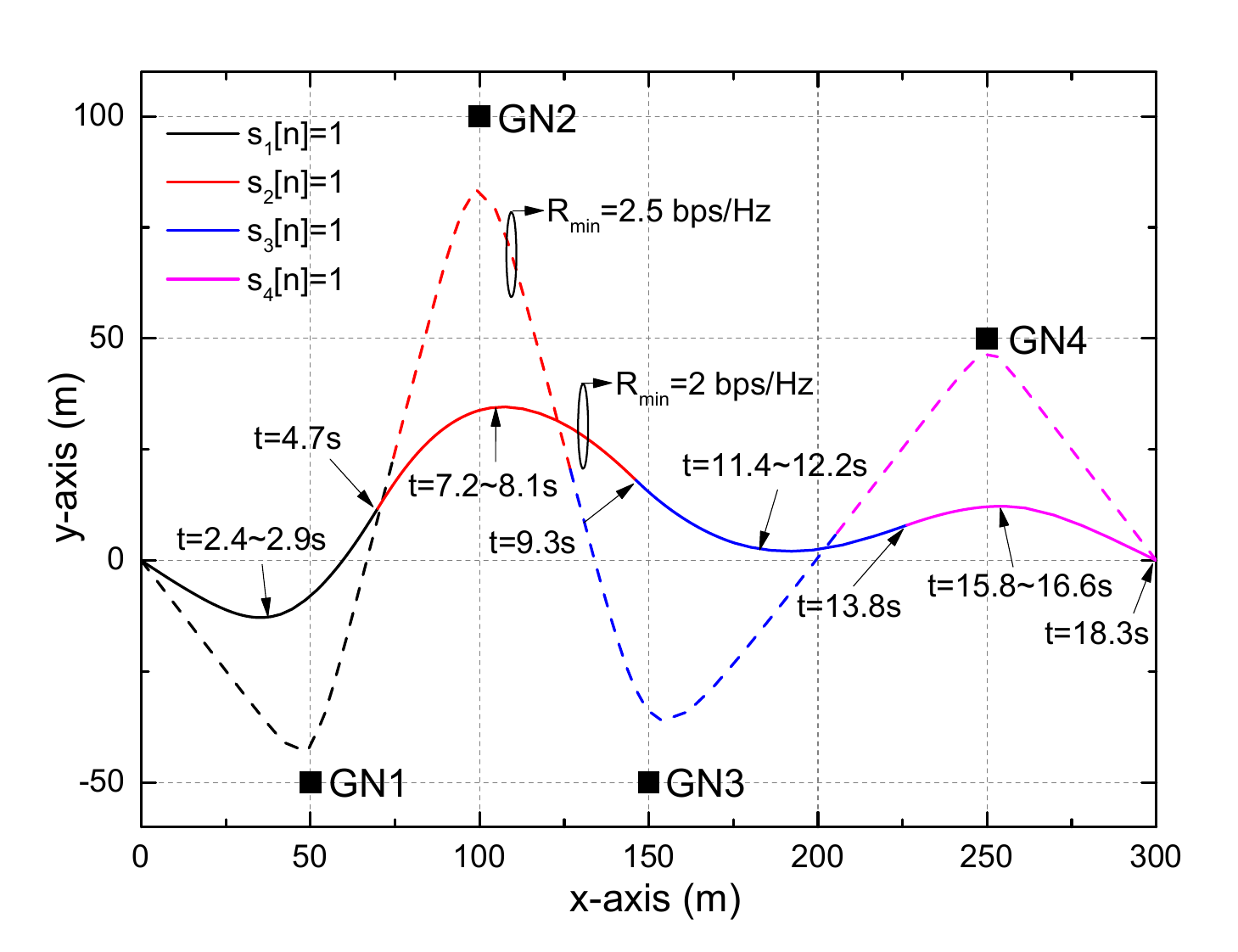}\label{R1-4}
    }
    \subfigure[Scheduling indicator.]{
      \includegraphics[width=0.315\linewidth]{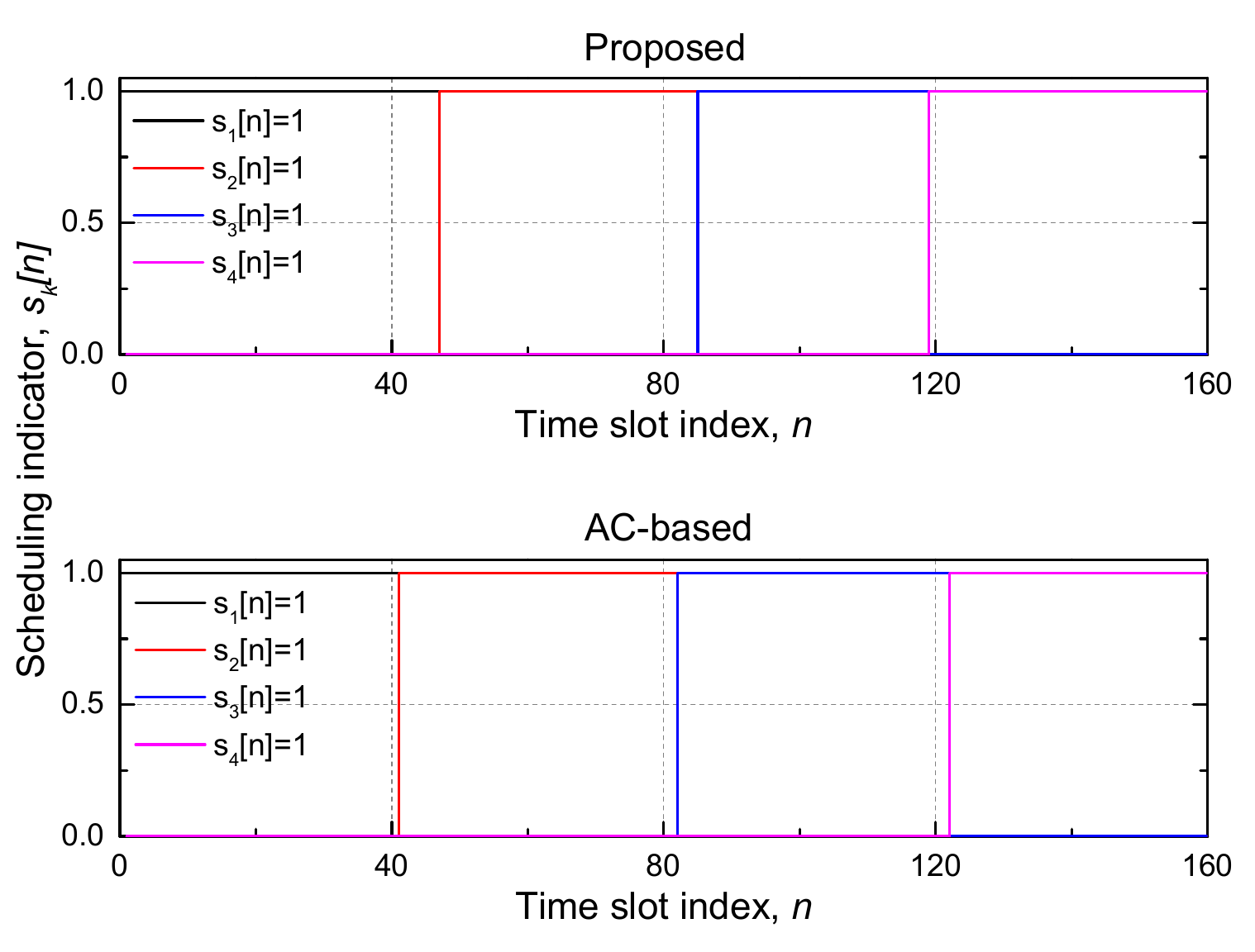}\label{R1-5}
    }
    \subfigure[Time slot length.]{
      \includegraphics[width=0.315\linewidth]{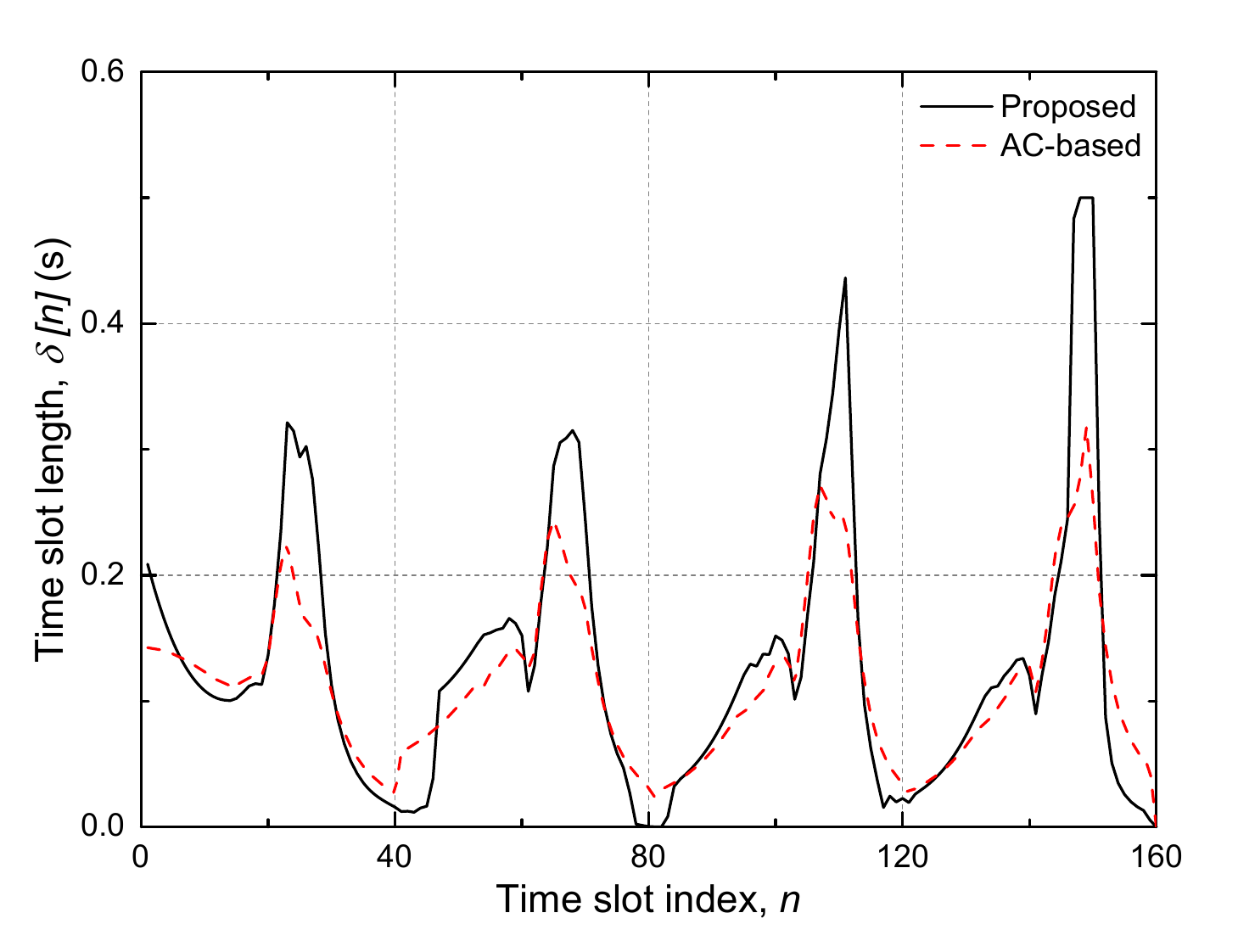}\label{R1-6}
    }
    \subfigure[LoS probability.]{
      \includegraphics[width=0.315\linewidth]{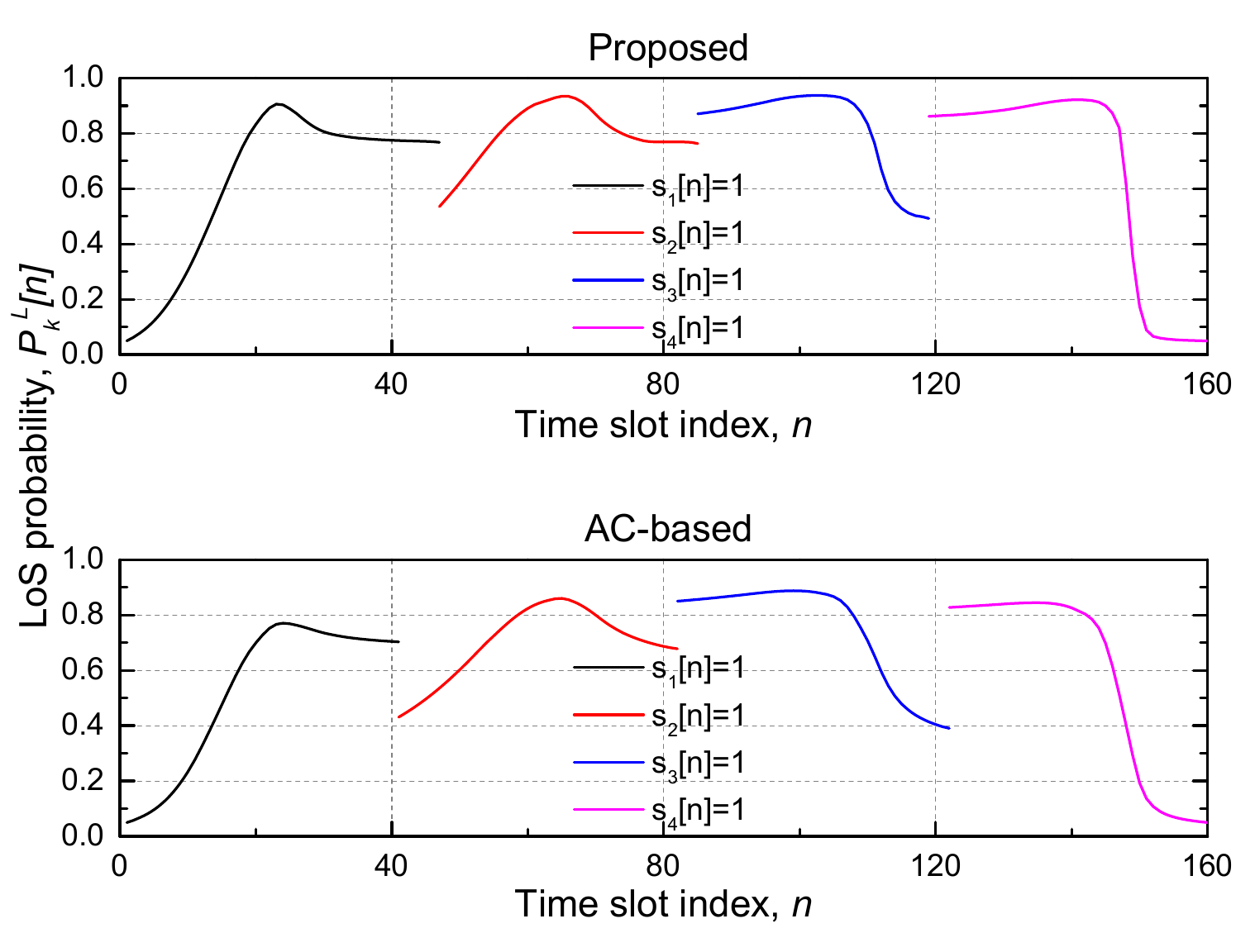}\label{R1-7}
    }
    \subfigure[Average SE of GNs.]{
      \includegraphics[width=0.315\linewidth]{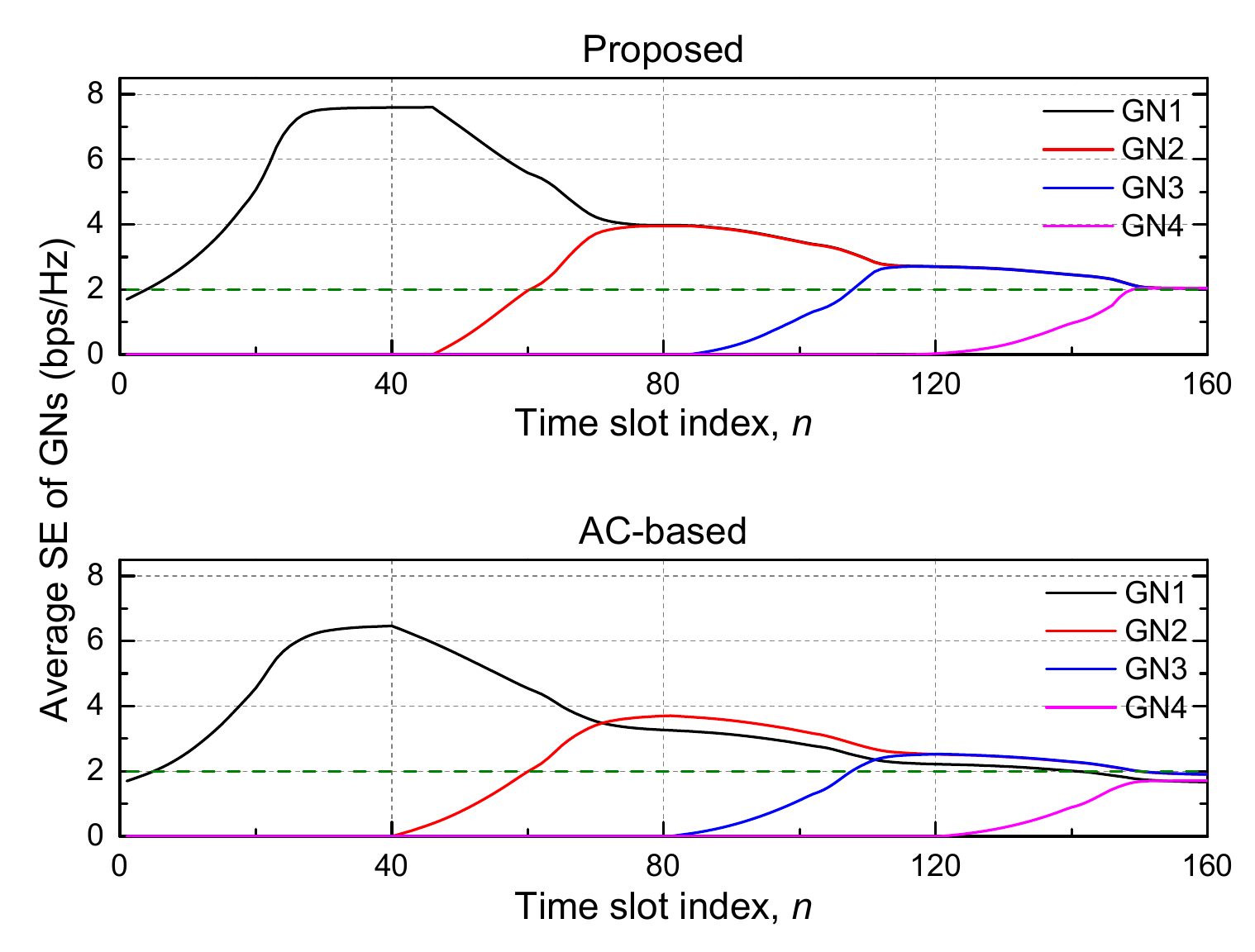}\label{R1-8}
    }
    \subfigure[Convergence behavior.]{
      \includegraphics[width=0.315\linewidth]{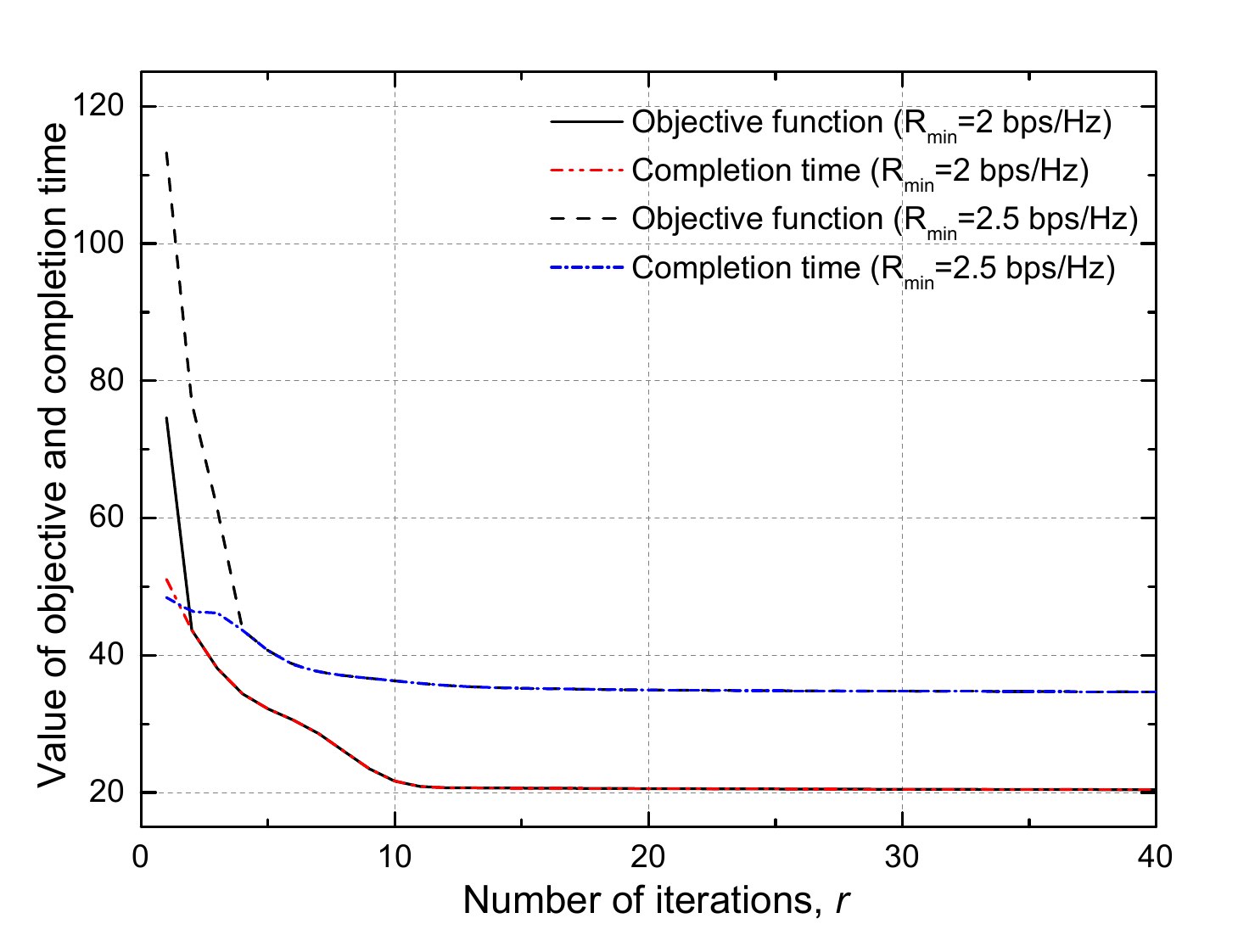}\label{R1-9}
    }
  \end{center}
\caption{Comparison of trajectory and resource allocation between the proposed and AC-based schemes, and convergence behavior.}
\label{R1}
\end{figure*}

The simulation parameters listed in Table~\ref{table1} are chosen to reflect representative settings widely adopted in UAV communication systems \cite{Zeng19,Meng22,Lei25,Pan22,Liu22,You20,Duo20,Luo21,He23,Park26,Kim25}. To assess the robustness of the proposed scheme against channel randomness, we conduct Monte Carlo simulations using the original stochastic channel model. For each optimized UAV trajectory, slot-time duration, and scheduling policy, independent realizations of small-scale fading and shadowing are generated in accordance with the assumed statistical distributions. The corresponding instantaneous uplink SE is computed for each realization, and the ergodic SE is estimated by averaging over $30{,}000$ Monte Carlo realizations. In addition, the following five schemes are considered for performance comparison. 
\begin{enumerate}
    
    \item \textit{Proposed scheme:} The UAV strategy $(\mathbf{S}, \mathbf{Q}, \boldsymbol{\Delta})$ is optimized using Algorithm~\ref{Alg1}.

    \item \textit{Average-channel-based (AC-based) scheme \cite{You20,Duo20}:} The UAV strategy is optimized based on the average-channel-based SE approximation in Section III-A. To compensate for potential infeasibility caused by SE overestimation, a positive margin is iteratively added to the target SE in the average-channel-based optimization until the resulting design satisfies the original minimum-SE requirement $R_{\textrm{min}}$ under the actual channel model. The margin step size is set to $10^{-4}$.

    \item \textit{Fixed-slot-length scheme:} Instead of optimizing the time-slot lengths individually, a single common slot length $\delta$ is optimized, resulting in a mission completion time of $T \!=\! N\delta$. Consequently, the UAV strategy $(\mathbf{S}, \mathbf{Q}, \delta)$ is optimized under the same system constraints.
    
    \item \textit{Fixed-altitude scheme:} The UAV altitude is fixed at $H_{\textrm{min}}$, and the UAV strategy, including $\mathbf{S}$, $\boldsymbol{\Delta}$, and the horizontal trajectory, is optimized.
    
    \item \textit{Fixed-trajectory scheme:} The UAV follows a hover-and-fly trajectory at an altitude $H_{\textrm{min}}$, sequentially hovering at each GN location and traveling in straight lines at the maximum velocity between GNs, while optimizing the remaining variables $\mathbf{S}$ and $\boldsymbol{\Delta}$.

\end{enumerate}
While the comparison with the AC-based scheme reveals the inherent limitations of average-channel-based SE approximations, the other baseline schemes assess the effect of successively restricting individual optimization variables.

Fig. \ref{R1} compares the UAV trajectory and resource allocation obtained by the proposed scheme and the AC-based scheme.
To clearly expose the limitation of average-channel-based SE approximations, the AC-based results are shown without applying any margin adjustment, so that the violation of the minimum-SE constraint can be directly observed.

\begin{figure*}[ht!]
  \begin{center}
    \subfigure[Completion time.]{
      \includegraphics[width=0.315\linewidth]{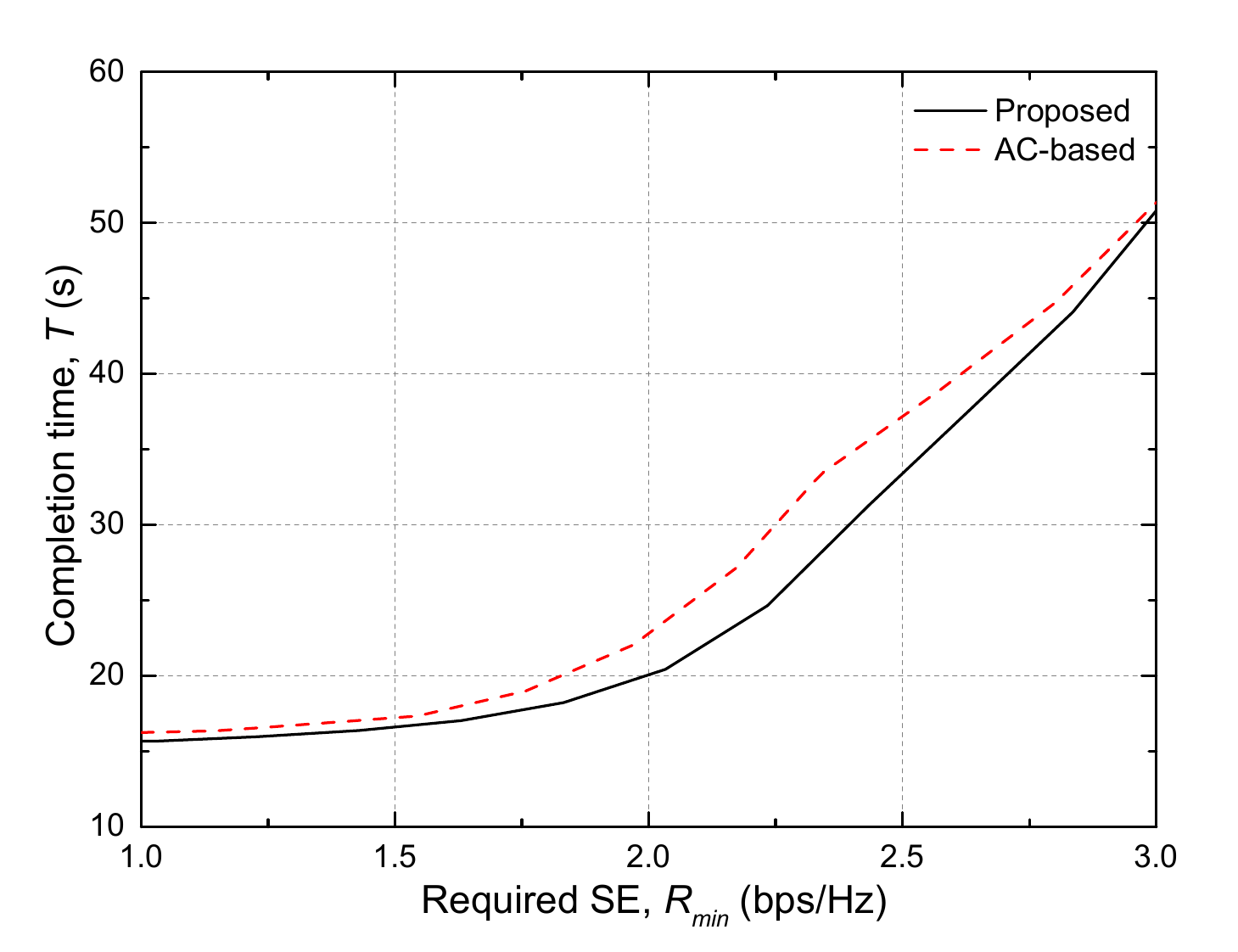}\label{R2-1}
    }
    \subfigure[Overestimated region.]{
      \includegraphics[width=0.315\linewidth]{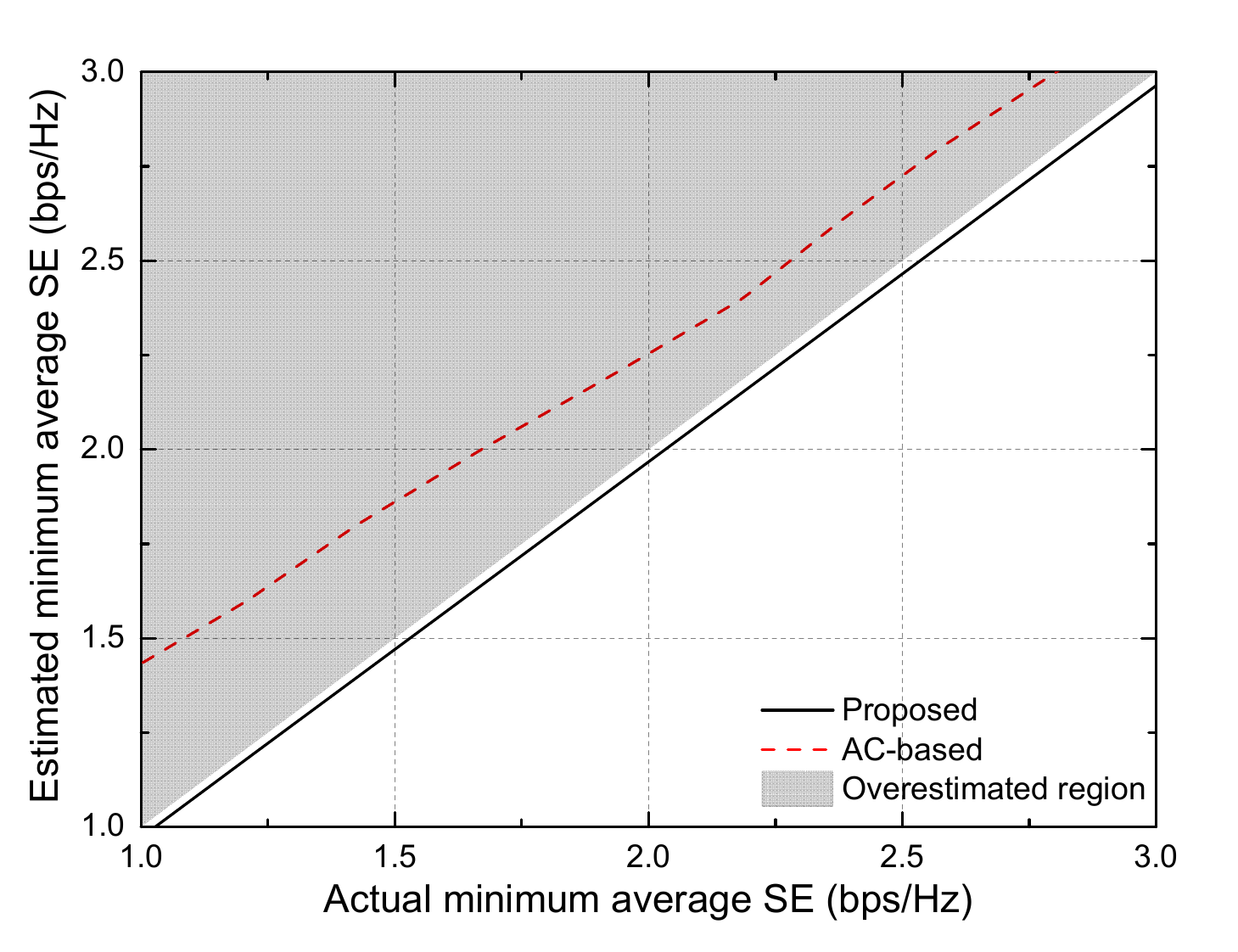}\label{R2-2}
    }
    \subfigure[Computation time.]{
      \includegraphics[width=0.315\linewidth]{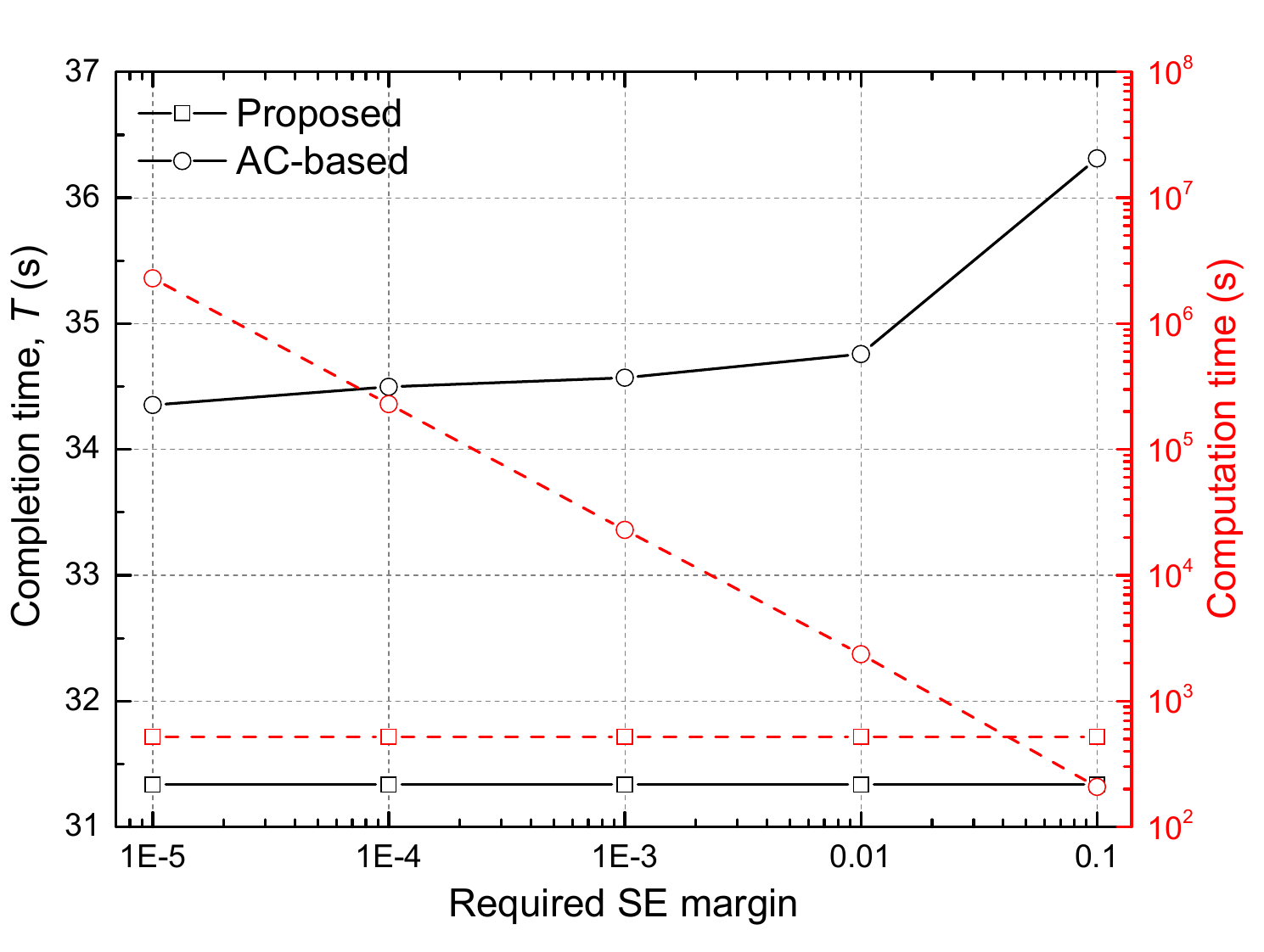}\label{R2-3}
    }
  \end{center}
\caption{Performance comparison between the proposed and AC-based schemes.}
\label{R2}
\end{figure*}

As shown in Fig. \ref{R1-1} and \ref{R1-2}, the UAV trajectory adapts noticeably to the required SE level $R_{\textrm{min}}$. When the required SE is high (e.g., $R_{\textrm{min}}$ = 2.5 bps/Hz), the UAV approaches each GN more closely and maintains a relatively high altitude, particularly while traveling between GNs. This behavior increases the elevation angle and, consequently, the LoS probability, enabling reliable rate provisioning under stringent SE requirements. In contrast, for a lower requirement (e.g., $R_{\textrm{min}}$ = 2 bps/Hz), the UAV prioritizes reducing the mission completion time by shortening the horizontal travel distance and limiting close approaches to each GN. To still satisfy the minimum-SE requirement, it compensates by flying at a higher altitude, which increases the elevation angle and the LoS probability, thereby maintaining reliable rate provisioning with reduced horizontal maneuvering. In these figures, the color-coded trajectory segments indicate the scheduled GN at each time slot, illustrating how the scheduling decisions are coordinated with the UAV trajectory.

In contrast, Fig. \ref{R1-3} and \ref{R1-4} show that the AC-based scheme yields a trajectory with a consistently lower altitude and a larger separation distance from each GN, compared with the proposed expected-SE-based design. This behavior stems from the average-channel-based SE approximation, which overestimates the true expected SE and thus enforces the minimum-SE constraint in an overly optimistic manner. As a result, the UAV does not need to sufficiently approach each GN or increase its altitude to enhance the LoS probability, leading to trajectories that remain lower and farther away from the GNs. As further confirmed in Fig. \ref{R1-5}, the corresponding scheduling indicators take binary values of 0 or 1, demonstrating that a valid TDMA-based scheduling policy is obtained and that each time slot is assigned to at most one GN.

As shown in Fig. \ref{R1-6}, the proposed scheme assigns longer time-slot durations when the UAV hovers above each GN, where the channel conditions are favorable, while shorter slots are allocated during inter-GN travel or scheduling transitions. Consistent with this behavior, by jointly examining Fig. \ref{R1-6} and Fig. \ref{R1-7}, it can be observed that longer time-slot durations tend to be assigned to time slots with higher LoS probability. In contrast, the AC-based scheme determines the time-slot lengths based on an average-channel-based SE estimate, which overestimates the achievable SE. Consequently, shorter slot durations are allocated even when the UAV is closest to each GN. Moreover, the AC-based scheme operates with a lower LoS probability than the proposed scheme, especially during scheduling transitions, since it does not sufficiently increase the altitude or reduce the service distance to enhance the LoS condition. As a result, a larger portion of transmission is performed under less favorable channel conditions.

These differences are directly reflected in the achieved average SE of GNs, as shown in Fig. \ref{R1-8}. While the proposed scheme satisfies the minimum-SE requirement precisely at the mission completion time for all GNs, the AC-based scheme fails to meet the target $R_{\textrm{min}}$ due to its overly optimistic SE estimation. This result confirms that the proposed scheme yields solutions that are feasible with respect to the original minimum-SE constraint, as it explicitly accounts for the true expected SE through a conservative and accurate SE estimation. In contrast, the AC-based scheme produces infeasible solutions that violate the minimum-SE requirement due to the overestimation inherent in the average-channel-based SE approximation.

Fig. \ref{R1-9} illustrates the convergence behavior of the proposed algorithm for $R_{\textrm{min}}$ = 2 and $R_{\textrm{min}}$ = 2.5 bps/Hz, showing both the objective value including the penalty term and the actual mission completion time. In the early iterations, the optimization variables are not sufficiently refined, making it difficult to satisfy the minimum-SE constraint. As a result, the penalty term becomes active, causing the objective value to exceed the completion time. As the optimization proceeds, the minimum-SE constraint is gradually satisfied and the penalty term diminishes. Consequently, the objective value converges to the mission completion time, indicating that the constraint violation has been eliminated. For both SE requirements, stable convergence is achieved within approximately 20 iterations.

Fig. \ref{R2} provides a comprehensive comparison between the proposed scheme and the AC-based scheme, highlighting the fundamental limitations of average-channel-based SE approximations. In the AC-based scheme, SE overestimation may lead to violations of the minimum-SE requirement, and thus the target $R_{\textrm{min}}$ is iteratively increased by adding a margin until the resulting design satisfies the minimum-SE constraint.

Fig. \ref{R2-1} compares the mission completion time of the proposed and AC-based schemes under different minimum-SE requirements $R_{\textrm{min}}$. As shown in the figure, the proposed scheme consistently achieves a shorter completion time than the AC-based scheme for all considered values of $R_{\textrm{min}}$. The performance gap between the two schemes is most pronounced in the intermediate regime of $1.75 \!\le\! R_{\textrm{min}} \!\le\! 2.75$ bps/Hz, where the minimum-SE requirement is neither overly stringent nor overly relaxed. In this intermediate regime, although NLoS transmissions can still be exploited from a mission completion time minimization perspective, accurately identifying when LoS conditions must be enforced becomes critical for performance. The proposed scheme correctly enforces LoS conditions in those critical segments where NLoS transmissions are insufficient to meet the minimum-SE requirement, while allowing NLoS transmissions elsewhere. In contrast, due to the systematic overestimation of achievable SE under NLoS channels, the AC-based scheme continues to rely on NLoS transmissions even in segments where LoS conditions are actually required, leading to mismatched trajectory and resource allocation decisions and a pronounced increase in mission completion time. When $R_{\textrm{min}} \!<\! 1.75$ bps/Hz, both schemes can easily satisfy the SE requirement. In this case, the UAV follows an almost straight-line trajectory from the initial to the final location to minimize the completion time, and transmissions are predominantly performed under NLoS conditions, yielding only a marginal performance gap. Conversely, when $R_{\textrm{min}} \!>\! 2.75$ bps/Hz, satisfying the SE requirement becomes challenging for both schemes, forcing most transmissions to be carried out directly above each GN. Consequently, LoS channels dominate, and the performance gap between the two schemes diminishes again in the high-$R_{\textrm{min}}$ regime.

Fig. \ref{R2-2} compares the SE estimated in the optimization with the actual SE obtained via Monte Carlo simulations for both schemes. For the AC-based scheme, all operating points lie in the overestimation region, indicating that the achievable SE is consistently overestimated due to the use of average-channel-based SE approximations. In contrast, the proposed scheme optimizes based on a conservative lower bound of the expected SE, resulting in estimated SE values that are consistently lower than the corresponding actual SE across all cases. This conservative property ensures that the minimum-SE constraint is satisfied with respect to the true expected SE, thereby guaranteeing the feasibility of the proposed scheme under the original problem formulation.

Fig. \ref{R2-3} compares the mission completion time and the computation time of the proposed and AC-based schemes as functions of the SE margin used in the AC-based scheme to satisfy the minimum-SE constraint. The proposed scheme does not require any margin adjustment and therefore shows identical completion time and computation time for all margin values. For the AC-based scheme, when a large margin (e.g., 0.1) is applied, the optimization requires fewer repetitions and thus results in a shorter computation time than the proposed scheme. However, such a large margin forces the AC-based optimization to target an SE level higher than $R_{\textrm{min}}$ to compensate for SE overestimation, which yields an overly conservative design and increases the completion time by over 15\% compared to the proposed scheme. As the margin decreases, the completion time of the AC-based scheme gradually decreases as well. Nevertheless, it remains consistently longer than that of the proposed scheme. Moreover, smaller margins necessitate repeated re-optimizations to eliminate violations of the minimum-SE constraint, which significantly increases the computation time compared to the proposed scheme. These results demonstrate that margin-based correction in the AC-based scheme introduces an unfavorable trade-off between performance and computational complexity, and cannot outperform the proposed scheme, which directly optimizes the trajectory and resource allocation based on an accurate characterization of the expected SE.

\begin{figure}[t!]
  \begin{center}
    \subfigure[$T$ vs. $V_{\textrm{max}}$.]{
      \includegraphics[width=0.9\linewidth]{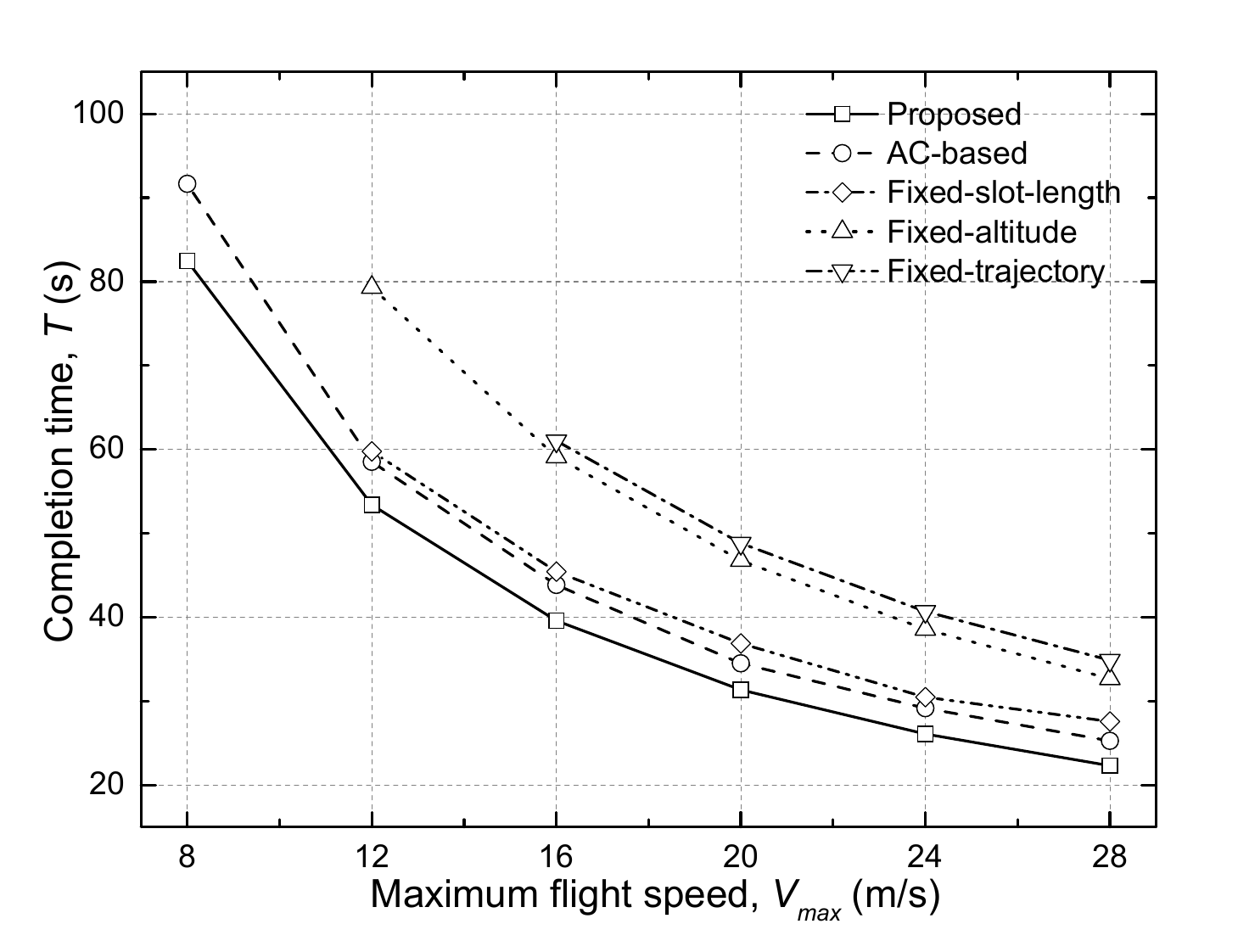}\label{R3-1}
    }
    \subfigure[$T$ vs. $K_R$.]{
      \includegraphics[width=0.9\linewidth]{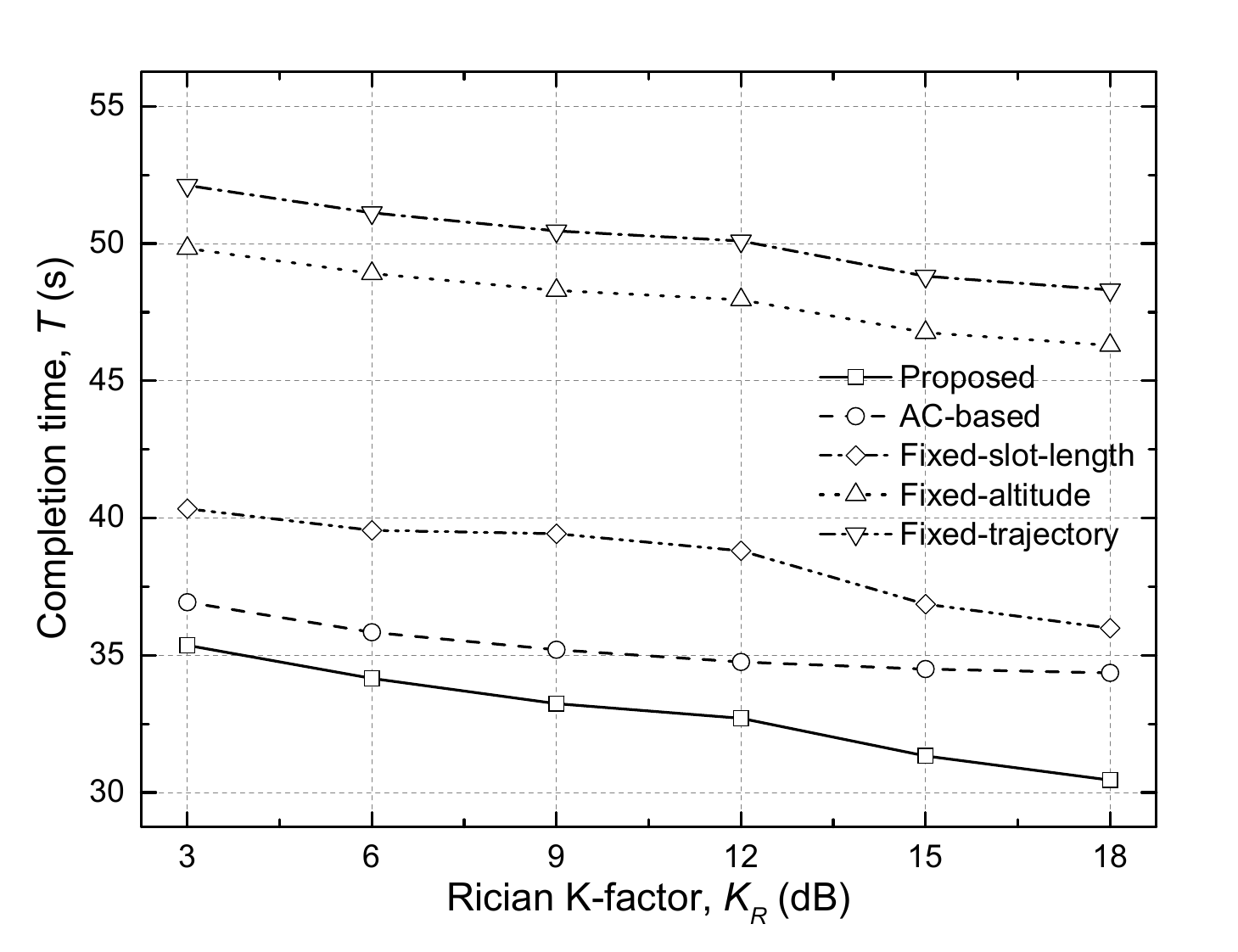}\label{R3-2}
    }
  \end{center}
\caption{Performance comparison between the proposed and baseline schemes.}
\label{R3}
\end{figure}

Fig. \ref{R3} presents the performance comparison between the proposed and baseline schemes under different system parameters.
Specifically, Fig. \ref{R3-1} compares the mission completion time with respect to the UAV maximum velocity ($V_{\textrm{max}}$). As $V_{\textrm{max}}$ increases, the UAV can travel faster between GNs, allowing it to spend a larger portion of time hovering above each GN for data collection. This increases the LoS probability and improves the channel quality, enabling the minimum-SE requirement $R_{\textrm{min}}$ to be satisfied more efficiently. As a result, all schemes exhibit a decreasing completion time as $V_{\textrm{max}}$ increases. The fixed-trajectory scheme becomes infeasible when $V_{\textrm{max}} \!<\! 16$ m/s. This is because it follows a predetermined hover-and-fly trajectory at the minimum altitude $H_{\textrm{min}}$, directly visiting and serving all GNs without allowing trajectory or altitude adaptation. Under a low UAV speed, the prolonged inter-GN travel time at low altitude results in insufficient SE accumulation, making it impossible to satisfy the minimum-SE requirement regardless of the mission duration. The fixed-altitude and fixed-slot-length schemes fail to satisfy $R_{\textrm{min}}$ when $V_{\textrm{max}} \!<\! 12$ m/s. In the fixed-altitude scheme, operating at the low altitude $H_{\textrm{min}}$ significantly reduces the LoS probability during inter-GN travel, resulting in a high proportion of NLoS transmission. As $V_{\textrm{max}}$ decreases, the time spent in these NLoS-dominated regions increases, resulting in inefficient SE and eventually making it impossible to satisfy the minimum-SE constraint. The fixed-slot-length scheme, on the other hand, can leverage altitude adaptation to achieve relatively higher LoS probability during inter-GN travel. However, since a common slot length is applied to all time slots, increasing the slot duration to enable hovering above GNs simultaneously increases the slot duration during inter-GN travel. Similar to the fixed-altitude case, a small $V_{\textrm{max}}$ leads to excessive SE loss during inter-GN travel due to prolonged NLoS transmission, rendering the minimum-SE constraint infeasible. Although the AC-based scheme performs optimization based on an inaccurate SE approximation, it allows flexible adjustment of time-slot durations across slots as well as the 3D trajectory. This flexibility enables the AC-based scheme to outperform the fixed-slot-length scheme by allocating shorter slots during unfavorable channel conditions, thereby highlighting the importance of time-slot length optimization. Nevertheless, due to SE overestimation, its performance remains inferior to that of the proposed scheme.

Fig. \ref{R3-2} shows the performance comparison with respect to the Rician $K$-factor ($K_R$). As $K_R$ increases, the LoS component of the wireless channel becomes more dominant, which reduces channel variability and increases the expected SE. Consequently, the mission completion time decreases for all schemes. For the AC-based scheme, however, the overestimation of the achievable SE leads to a tendency to minimize trajectory movement and rely more heavily on transmission under NLoS conditions in order to reduce the completion time. As a result, the AC-based scheme fails to fully exploit the strengthened LoS component as $K_R$ increases, causing the performance gap between the proposed and AC-based schemes to become more pronounced. Since the baseline schemes achieve performance in the order of the AC-based, fixed-slot-length, fixed-altitude, and fixed-trajectory schemes, this result indicates that system performance is increasingly influenced by flexibility in 3D trajectory design, followed by altitude optimization, time-slot length adaptation, and accurate SE approximation, in that order. Overall, the proposed scheme consistently outperforms all baseline schemes across all values of the system parameters, demonstrating the effectiveness of jointly optimizing the user scheduling, UAV 3D trajectory, time-slot durations, and the proposed effective SE approximation.

\section{Conclusions}

This paper addressed the mission completion time minimization problem in UAV-assisted wireless communication systems under minimum expected SE constraints. Departing from conventional average-channel-based designs, we proposed a conservative and tractable expected-SE formulation that explicitly accounts for the stochastic nature of probabilistic LoS propagation, small-scale fading, and shadowing. By constructing a computable lower bound via CDF-domain discretization and quadrature-based reformulation, the proposed approach enables reliable constraint enforcement while remaining amenable to joint trajectory and resource optimization. To solve the resulting mixed-integer nonconvex problem, we developed a penalty-based block coordinate descent framework that alternately optimizes user scheduling and the UAV trajectory together with adaptive time-slot durations. The SCA and quadratic transform techniques were employed to handle the coupled nonconvexities, ensuring feasibility with respect to the original expected-SE constraints while achieving efficient convergence with polynomial-time complexity. Beyond performance improvements, the results clearly show that average-channel-based approaches frequently yield infeasible solutions, as overestimation of the expected SE causes violations of the minimum SE requirements under the true stochastic channel conditions. In contrast, the proposed approach derives a conservative and accurate expected-SE lower bound that guarantees feasibility and ensures stable performance across all operating regimes. Consequently, the proposed framework not only shortens the mission completion time but also provides a robust and practically deployable design methodology for UAV communication systems based on accurate expected-SE modeling under probabilistic LoS channels.


\end{document}